\documentclass[twocolumn,apj]{aastex631}

\let\splitbox\relax
\usepackage{adjustbox}
\usepackage{amsmath}

\begin{document}

\title{A parametric study of solar wind properties and composition using fluid and kinetic solar wind models}

\correspondingauthor{Alexis P. Rouillard}
\email{arouillard@irap.omp.eu}

\author[0009-0002-9442-5191]{Paul Lomazzi}
\affiliation{Universit\'e de Toulouse \\
UPS-OMP; IRAP; Toulouse, France CNRS \\
IRAP; 9 Av. colonel Roche, BP 44346}

\author[0000-0003-4039-5767]{Alexis P. Rouillard}
\affiliation{Universit\'e de Toulouse \\
UPS-OMP; IRAP; Toulouse, France CNRS \\
IRAP; 9 Av. colonel Roche, BP 44346}

\author{Michael A. Lavarra}
\affiliation{Universit\'e de Toulouse \\
UPS-OMP; IRAP; Toulouse, France CNRS \\
IRAP; 9 Av. colonel Roche, BP 44346}

\author[0000-0002-1814-4673]{Nicolas Poirier}
\affiliation{Rosseland Centre for Solar Physics, University of Oslo, P.O. Box 1029, Blindern, NO-0315 Oslo, Norway}

\author{Pierre-Louis Blelly}
\affiliation{Universit\'e de Toulouse \\
UPS-OMP; IRAP; Toulouse, France CNRS \\
IRAP; 9 Av. colonel Roche, BP 44346}

\author[0000-0002-1628-0276]{Jean-Baptiste Dakeyo}
\affiliation{Space Sciences Laboratory, University of California, Berkeley, CA, USA}

\author[0000-0001-5014-7682]{Viviane Pierrard}
\affiliation{Royal Belgian Institute for Space Aeronomy,
Space Physics and STCE,
3 av. circulaire, B-1180 Brussels, Belgium}
\affiliation{Universit\'e Catholique de Louvain, ELI-C,
3 place Louis Pasteur, B-1348 Louvain-la-Neuve, Belgium}

\author[0000-0002-2916-3837]{Victor Réville}
\affiliation{Universit\'e de Toulouse \\
UPS-OMP; IRAP; Toulouse, France CNRS \\
IRAP; 9 Av. colonel Roche, BP 44346}

\author[0000-0001-8583-8619]{Christian Vocks}
\affiliation{Leibniz-Institut für Astrophysik Potsdam (AIP), An der Sternwarte 16, 14482 Potsdam, Germany}

\author[0000-0001-7119-9545]{Simon Thomas}
\affiliation{Universit\'e de Toulouse \\
UPS-OMP; IRAP; Toulouse, France CNRS \\
IRAP; 9 Av. colonel Roche, BP 44346}

\begin{abstract}

The physical processes in the solar corona that shape the solar wind remain an active research topic. Modeling efforts have shown that energy and plasma exchanges near the transition region plays a crucial role in modulating solar wind properties. Although these regions cannot be measured in situ, plasma parameters can be inferred from coronal spectroscopy and ionization states of heavy ions, which remain unchanged as they escape the corona. We introduce a new solar wind model extending from the chromosphere to the inner heliosphere, capturing thermodynamic coupling across atmospheric layers. By including neutral and charged particle interactions, we model the transport and ionisation processes of the gas through the transition region, the corona and into the solar wind. Instead of explicitly modeling coronal heating, we link its spatial distribution to large-scale magnetic field properties. Our results confirm that energy deposition strongly affects wind properties through key mechanisms involving chromospheric evaporation, thermal expansion, and magnetic flux expansion. For sources near active regions, the model predicts significant solar wind acceleration, with plasma outflows comparable to those inferred from coronal spectroscopy. For winds from large coronal holes, the model reproduces the observed anticorrelation between charge state and wind speed. However, the predicted charge state ratios are overall lower than observed. Inclusion of a population of energetic electrons enhances both heavy ion charge states and solar wind acceleration, improving agreement with observations.

\end{abstract}



\section{Introduction}

The mechanisms in the corona that control the bulk properties of the solar wind are still under debate. Previous theoretical studies have provided evidence that energy and plasma flows through the transition region play a central role in the mass flux escaping the solar corona \citep[e.g.][]{hansteen_coronal_1995}. These energy exchanges in turn depend on the heating processes at work in the lower solar atmosphere, which are poorly understood.  \\

The source region of the solar wind plasma is still collisional in the corona, transitioning to a collisionless state at $\sim 1$ solar radius above the photosphere. In the collisional region, heavy ions undergo collisions with hot electrons and become increasingly ionised with height, and this until electron densities are so low and their temperatures so high that collisional processes vanish and the charge state of an ion can no longer change and becomes 'frozen-in' \citep{Geiss_1995}.\\

Ion charge states measured remotely or in-situ therefore provide information on solar wind source conditions and, in particular, on electron source temperatures.  Previous observational studies using data from Ulysses and from the NASA Advanced Composition Explorer (ACE) \citep{stone_ace_1998}
have shown that the charge state in fast and slow solar wind streams differ greatly \citep{Geiss_1995,zhao_global_2009}. In particular, they showed that the charge state ratios as $O_7^+/O_6^+$ and $C_6^+/C_5^+$ tend to be higher in slow winds than in fast winds, suggesting a hotter coronal origin of the slow wind. \\

While coronal holes are widely accepted as the source region of the Alfvénic fast wind, the source location of slow wind is still under debate and is thought to have multiple source components including dynamic solar streamers \citep[e.g.][]{rouillard_multispacecraft_2009,rouillard_observations_2010}, boundaries of coronal holes \citep{wang_slow_2009} and low-latitude coronal holes \citep{damicis_origin_2015,wang_observations_2019}. The temperatures of the dynamic sources of the slow solar wind originating in streamers are complex to grasp because they involve periodic plasma exchanges between loops and the open field \citep[e.g.][]{reville_flux_2022}. \\

Assuming that the fast and slow wind components originate in the more quiescent coronal holes, previous theoretical studies based on fluid wind models have tried to reproduce the anticorrelation between wind speed and charge state \citep{cranmer_self-consistent_2007,cranmer_suprathermal_2014,oran_steady-state_2015}. Although they have been successful in obtaining an anticorrelation, the resulting charge states are orders of magnitude lower than those measured in-situ. \\

The sources of these discrepancies are still unknown. \cite{ko_empirical_1997} and \cite{esser_differential_2001} proposed that differential flows between minor and major species could lead to increases in the charge state. A extra population of ionizing suprathermal electrons was also invoked to explain the ionization rates of minor ions \citep{ko_limitations_1996,esser_reconciling_2000,oran_steady-state_2015,cranmer_suprathermal_2014}. \\

An accurate model of the ion charge states in the solar wind requires not only a detailed transport model of thermal particles accounting for the thermodynamic coupling of the different layers of the low solar atmosphere but also of the transport of non-thermal particles and their interactions with the thermal population. This is a challenging task that ideally should account for the detailed evolution of the particle distribution function of all species.\\   

In the present paper, we use a more tractable approach by using the IRAP Solar Atmospheric Model ISAM developed in \cite{lavarra_2022} in combination with an exospheric kinetic solar wind model \citep{pierrard_exospheric_2023} to study the different processes that could influence the ionisation states of heavy ions in the different types of solar winds.  This approach allows us to model the ionising effects of energy deposition, non-thermal particles and differential flows for different conditions at the sources of the fast and slow solar winds. The model is constrained by several observations, including the SOlar and Heliospheric Observatory \citep[SoHO][]{domingo_soho_1995} spectroscopy and in-situ data from the Solar Orbiter \citep{muller_solar_2020} and Parker Solar Probe \citep{kinnison_parker_2020} missions. \\

This paper is organized as follows: after a presentation of the models used in section \ref{sec:num}, we begin by solving the transport of neutral and charged particles from the upper chromosphere to the solar wind in sections \ref{sec:HL} \& \ref{sec:sim_ar}. For this, we use the results of the recent coronal-interplanetary wind connectivity study of \cite{dakeyo_testing_2024} to specify the input parameters necessary to model the fast and slow solar winds in Section \ref{sec:solo}. This provides first estimates of the charge state of ions for different coronal-wind conditions. In section \ref{sec:solu_supra}, the output of the thermal model is then used as input to an exospheric solar wind model which provides the radial evolution of non-thermal electrons. This allows us to model the additional ionizing effects on ions of these non-thermal electrons. 

\section{Description of the Numerical Model}
\label{sec:num}
\subsection{The Conservation Equations}

To model the thermal component of the coupled chromosphere-corona-wind system, we use ISAM which dynamically solves the transport and interactions of neutrals and charged particles. The model uses a 16-moment set of transport equations described in detail in \cite{blelly_comparative_1993}. A first version of this model, IPIM (IRAP Plasmasphere-Ionosphere Model), was developed to simulate plasma transport in Earth's upper atmosphere \citep{marchaudon_new_2015} and has been thoroughly checked against a broad range of terrestrial observations \citep{marchaudon_impact_2020} and adapted to other planetary atmospheres \citep{blelly_transplanet_2019}. It was later modified to model the solar atmosphere by solving neutrals together with charged particles to simulate the variable low solar atmosphere for open field configurations \citep{lavarra_2022} and magnetic loops \citep{aurelien_transition_2022}. \\

Other implementations of the 16-moment approach to model the solar atmosphere are detailed in \cite{li_16-moment_1999} and later \cite{lie-svendsen_16-moment_2001}, \cite{janse_solar_2007} and \cite{byhring_modeling_2011}. A comparison between ISAM and these earlier applications of the 16-moment models is discussed in \cite{lavarra_2022}.\\

The density $n_s$, bulk flow velocity $u_s$, parallel $T_s^{\parallel}$ and perpendicular temperatures $T_s^{\perp}$, parallel heat flux $q_s^{\parallel}$ and perpendicular heat fluxes $q_s^{\perp}$ are solved along a given magnetic flux tube through the following set of conservation equations: 
{
\small
   \begin{equation}
      \frac{\partial n_s}{\partial t}+u_s\nabla_\parallel n_s+\frac{n_s}{A}\nabla_\parallel (\,Au_s )\,=\frac{\delta n_s}{\delta t} \label{eq:N}
   \end{equation}
   \begin{equation}
       \begin{split}
      \frac{\partial u_s}{\partial t}+u_s\nabla_\parallel u_s+\frac{\nabla_\parallel n_s k_B T_s^{\parallel}}{n_sm_s}+\frac{k_B}{m_s}\left(T_s^{\parallel}-T_s^{\perp}\right)\frac{\nabla_\parallel A}{A}\\+\frac{GM_\odot}{r^2}\cos(\theta)-\frac{F_s}{n_sm_s}=\frac{\delta u_s}{\delta t} \label{eq:U}
       \end{split}
   \end{equation}
  \begin{equation}
       \begin{split}
      \frac{\partial T_s^{\parallel}}{\partial t}+u_s\nabla_\parallel T_s^{\parallel} +2T_s^{\parallel}\nabla_\parallel u_s + \frac{\nabla_\parallel q_s^{\parallel} }{n_sk_B} + \frac{q_s^{\parallel}-2q_s^{\perp}}{n_sk_B}\frac{\nabla_\parallel A}{A}\\=\frac{(Q_s^{\parallel}-\Lambda_s^{\parallel} )\,}{n_sk_B}+\frac{\delta T_s^{\parallel}}{\delta t}\label{eq:Tp}
       \end{split}
   \end{equation}
     \begin{equation}
       \begin{split}
      \frac{\partial T_s^{\perp}}{\partial t}+u_s\nabla_\parallel T_s^{\perp} + \frac{\nabla_\parallel q_s^{\perp} }{n_sk_B} + \left(\frac{2q_s^{\perp}}{n_sk_B}+u_sT_s^{\perp}\right)\frac{\nabla_\parallel A}{A}\\=\frac{(Q_s^{\perp}-\Lambda_s^{\perp} )\,}{n_sk_B} +\frac{\delta T_s^{\perp}}{\delta t} \label{eq:Tt}
       \end{split}
   \end{equation}
     \begin{equation}
       \begin{split}
      \frac{\partial q_s^{\parallel}}{\partial t}+u_s\nabla_\parallel q_s^{\parallel} + 4q_s^{\parallel}\nabla_\parallel u_s + \frac{3n_sk_B^2T_s^{\parallel}}{m_s}\nabla_\parallel T_s^{\parallel} \\+ u_s q_s^{\parallel}\frac{\nabla_\parallel A}{A}=\frac{\delta q_s^{\parallel}}{\delta t} \label{eq:qp}
       \end{split}
   \end{equation}
     \begin{align}
       \begin{split}
      \frac{\partial q_s^{\perp}}{\partial t}+u_s\nabla_\parallel q_s^{\perp} + 2q_s^{\perp}\nabla_\parallel u_s + \frac{n_sk_B^2T_s^{\parallel}}{m_s}\nabla_\parallel T_s^{\perp}\\ +\left( \frac{n_sk_B^2T_s^{\perp}}{m_s}\left(T_s^{\parallel}-T_s^{\perp}\right)+ 2u_s q_s^{\perp} \right)\frac{\nabla_\parallel A}{A}\\=\frac{\delta q_s^{\perp}}{\delta t} \label{eq:qt}
       \end{split}
     \end{align}
}

Here, the subscript $s$ indicates the species considered, $k_B$ the Boltzmann constant, $m_s$ the atomic mass of the species, $G$ the gravitational constant, $M_\odot$ the solar mass, $\theta$ the angle between the radial direction and the local direction of the magnetic field line along which the equations are solved, $\Lambda_s^{\parallel(\perp)}$ the perpendicular and parallel radiative cooling terms, $Q_s^{\parallel(\perp)}$ the perpendicular and parallel heating terms. The terms on the right-hand side of the equations containing $\delta/\delta t$ represent collisional effects. The parallel and perpendicular heat fluxes represent the transport along the field lines of the parallel and perpendicular energies.\\

For ions, $F_s$ corresponds to the electrostatic polarization field contribution:
   \begin{equation}
   \begin{split}
      \frac{F_s}{m_s n_s}=-\frac{Z_s}{m_s}\frac{1}{n_e}\left[\nabla_\parallel(n_ek_BT^\parallel_e)+k_B(T^\parallel_e-T^\perp_e)\frac{1}{A}\nabla_\parallel A\right]\\+Z_s\frac{m_e}{m_i}\frac{\delta u_e}{\delta t}
      \label{eq:F_s}
   \end{split}
   \end{equation}
where the subscript $s$ indicates the ion considered and $Z_s$ represents the number of elementary charges $e$ carried by the ion.

Equations (1) \& (2) are not solved for the electrons, instead we impose quasi-neutrality and no field-aligned current $J=0$ which allows us to have :
\begin{align}
n_e &= \sum_{s=\text{ions}} \left( Z_s n_s \right) \\
u_e &= \frac{\sum_{s=\text{ions}} \left( Z_s n_s u_s \right)}{n_e}
\end{align}

The total temperature and heat flux are related to their parallel and perpendicular component by:

   \begin{equation}
      T=(T_s^{\parallel}+2T_s^{\perp})/3
      \label{eq:Ttot}
   \end{equation}
   \begin{equation}
      q=(q_s^{\parallel}+2q_s^{\perp})/2
      \label{eq:qTot}
   \end{equation}

We begin the study by computing the coupled transport of electrons, protons and hydrogen atoms along magnetic flux tubes rooted in the chromosphere and extending outwards into the solar wind. \\

\subsection{The Numerical Scheme}
To solve the above coupled differential equations, ISAM uses the flux-corrected algorithm for solving generalised continuity equations (LCPFCT) \citep{boris_flux-corrected_1976,boris_lcpfct-flux-corrected_nodate}. This finite volume method is based on an explicit second order conservative Godunov scheme. The scheme is non-centered and stabilised up to fourth order thanks to a numerical antidiﬀusive stage, but in the case of steep gradients the scheme is degraded to ﬁrst order accuracy by a stabilisation procedure characterised by a diﬀusive numerical ﬂux. ISAM solves the conservative equations including transport and local source terms. The time dependence is solved using a fourth-order Runge-Kutta algorithm. The conservation equations \ref{eq:N}-\ref{eq:qt} are integrated between an inner boundary located in the chromosphere at $\sim 600$ km above the photosphere and all the way to 20 R$_\odot$.\\

\subsection{The Model Setup}

\subsubsection{The Open Magnetic Field Model}
The magnetic flux tube geometry along which ISAM solves the plasma transport equations is given by the widely form inferred by \cite{kopp_dynamics_1976} from the analysis of eclipse observations:

   \begin{equation}
      A=A_0\left(\frac{r}{R_\odot}\right)^2\left(\frac{f_{max}e^{(r-r_g)/\sigma_g}+f_g}{e^{(r-r_g)/\sigma_g}+1}\right)
      \label{eq:AA0}
   \end{equation}

where $A(r)$ is the section of the flux tube with $A_0=A(R_\odot)$. $r_g$ and $\sigma_g$ are parameters describing respectively the height where the expansion rate is the strongest and the steepness of the expansion. $f_{max}$ is the asymptotical value of $(A/A_0)/(R_\odot/r)^2$ describing the deviation from a radial expansion. $f_g$ is chosen to have $A(R_\odot)=A_0$ so $f_g=1+(1-f_{max})e^{(R_\odot-r_g)/\sigma_g}$. In this study, only the maximum expansion factor $f_{max}$ is changed and the other parameters are kept fixed with $\sigma_g=0.5R_\odot$ and $r_g=1.3R_\odot$. The values for $f_{max}$ will be given by a global magnetic model of the solar atmosphere in the context of a magnetic connectivity analysis of Solar Orbiter data. 

\subsubsection{The Atmospheric Heating Rates}
\label{sec:heating_rates}

The heating processes are still highly debated and could have multiple origins in the different layers of the solar atmosphere. To obtain solutions of thermal proton and electron properties as close as possible to a broad range of remote sensing observation. The heating terms $Q_s^{\parallel(\perp)}$ in equation (3) \& (4) are a sum of three different heating functions acting on the transition region heating $Q_{TR,s}^{\parallel(\perp)}$, the corona $Q_{cor,s}^{\parallel(\perp)}$ and the solar wind $Q_{sw,s}^{\parallel(\perp)}$. \\

\begin{figure}[!h]
      \centering
      \includegraphics[scale=0.215]{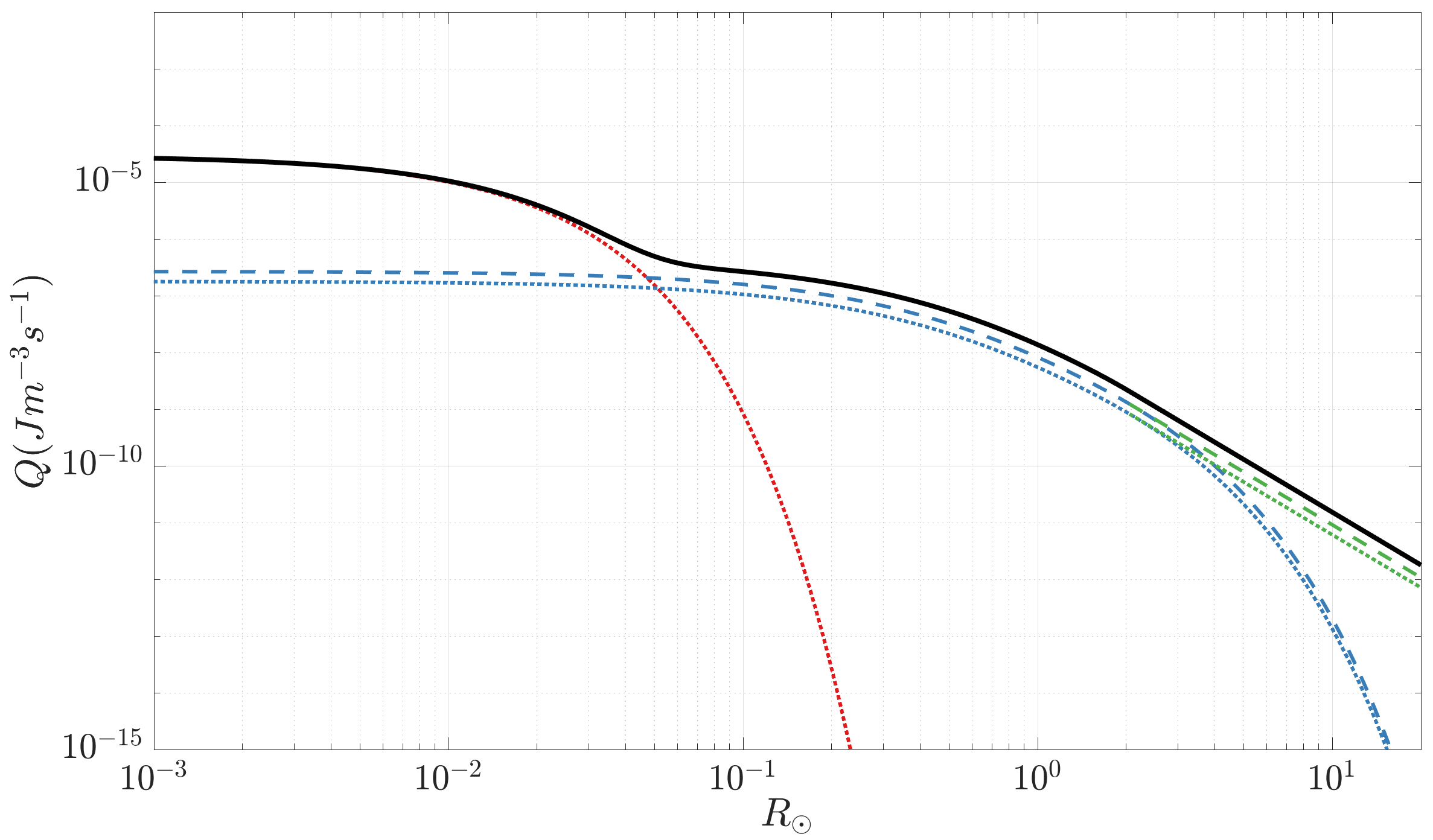}
      \caption{\emph{Heating function used for the fast wind solution from the model.\/}  Heating on the protons and electrons are indicated by using respectively dotted and dashed lines. Total heating (black) is the result of a transition region heating (red), a coronal heating (blue) and a solar wind heating (green).
      }
      \label{fig:heatingdist}
  \end{figure}

The heating repartition between directions and species is given in Figure \ref{fig:heatingdist}. For the transition region and coronal heating terms $Q_{TR,s}^{\parallel(\perp)}$ and $Q_{cor,s}^{\parallel(\perp)}$ we use the same phenomenological form as used in \cite{pinto_multiple_2017}, which was initially introduced by \cite{withbroe_temperature_1988}, \cite{mckenzie_acceleration_1995} and \cite{rifai_habbal_flow_1995}:

   \begin{equation}
      Q=-\nabla\cdot F_h
      \label{eq:qdiv}
   \end{equation}
where
   \begin{equation}
      F_h=F_{B_0} \frac{A_0}{A}e^{ -\frac{r-R_\odot}{H_f}}
      \label{eq:withbroe}
   \end{equation}
   
This functional form supposes a total energy flux $F_{B_0}$ proportional to $B_0$ and a heating scale height $H_f$ that controls the altitude range of energy deposition. As in \cite{wang_slow_2009}, this heating scale height is made inversely dependent of the maximum expansion factor $f_{max}$. Although the mechanisms heating the solar atmosphere are not certain, the dissipation of Alfvén waves channeled by the interaction of upward propagating and reflected counter-propagating waves reproduce a number of coronal properties \citep{verdini_alfven_2007,verdini_origin_2012,cranmer_self-consistent_2007}. In essence, the value of $f_{max}$ could define the radial gradients in Alfvén speed along a flux tube, which in turn should dictate the level of wave reflection and associated energy dissipation along that tube. A high value of $f_{max}$ is expected to lower the scale height of energy deposition by concentrating wave dissipation in the lower corona. \\

The following functional forms of the terms in $Q_{cor,s}^{\parallel(\perp)}$ and $Q_{TR,s}^{\parallel(\perp)}$ were found to reproduce well the typical height of the transition region and some general features of the solar corona:
\begin{equation}
    F^{cor}_{B_0}=0.86\times 10^5 \,B_0 \:\:\: erg \, cm^{-2} \, s^{-1}
    \label{eq:Fcor}
\end{equation}
\begin{equation}
    H^{cor}_{f}=4.35\times f_{ss}^{-0.75} \:\: R_\odot
    \label{eq:Hcor}
\end{equation}
\begin{equation}
    F^{TR}_{B_0}=0.43\times 10^5 \,B_0 \:\:\: erg \,cm^{-2}\,s^{-1}
    \label{eq:Ftr}
\end{equation}

$H^{TR}_f$ is kept unchanged in all runs and we chose $H^{TR}_f=0.01R_\odot$. The proportional relationship between the energy flux and $B_0$ was demonstrated in \citep{wang_2016} using OMNI data \citep{king_2005}, where the non-gravitational energy flux was derived from the mass and energy conservation equations.\\

When turbulence reaches dissipative scales, multiple kinetic mechanisms will heat the plasma and distribute the energy between the different species \citep{axford_origin_1992,kohl_uvcssoho_1998,maneva_dissipation_2015}. It is well established that ion and electron heating is sustained far out into the interplanetary medium \cite{marsch_kinetic_2006} as evidenced by the elevated electron temperatures measured beyond 1AU \cite{issautier_solar_1998} and the presence of $T_\perp > T_\parallel$ of ions beyond $0.3$AU.\\

The radial distribution of this interplanetary wind heating  $Q_{sw,s}^{\parallel(\perp)}$ is chosen here to follow a power law starting at the height $r_s^{sw}$ beyond which the coronal heating is replaced by the wind heating. The power law begins at $Q_{cor,s}^{\parallel(\perp)}$ at $r_s^{sw}$ with a fixed slope defined by the derivative of $Q_{cor,s}^{\parallel(\perp)}$ at that radial distance:

   \begin{equation}
      Q_{sw}= Q_{cor,s}^{\parallel(\perp)}(r_s^{sw}) \left(\frac{r(r_s^{sw}:20R\odot)}{r_s^{sw}}\right)^{(\Delta \log Q_{cor,s}^{\parallel(\perp)}/\Delta \log r)|_{r_s^{sw}}} 
      \label{eq:Qsw}
   \end{equation}
   
where $\Delta(*)$ indicates the difference of ($*$) between two successive grid points.\\

In this formulation, the amount of heating in the solar wind is only controlled by changing $r_s^{sw}$. This choice was made to keep a smooth transition between the coronal and solar wind heating whilst still benefiting from the exponential expression of the coronal heating term.\\

Heating is applied only to electrons in the transition region. This energy is transmitted to other particles through the frequent Coulomb collisions occurring in that region. Coronal and solar wind heating are applied $60\%$ to protons and $40\%$ to electrons to get, as we shall see, reasonable electron temperatures compared to spectroscopic observations. $80\%$ of the proton heating is applied in the perpendicular direction and $20\%$ in the parallel direction. The same ratio between parallel and perpendicular heating is applied to electrons. This heating repartition gave optimal result from many iterations tests and is kept for all simulations presented in this paper.\\


\subsection{Radiative Energy Loss:}

The radiative cooling term, $\Lambda$, is based on the optically thin function of \citep{athay_radiation_1986} with a correction factor to account for optical thickness in the upper chromosphere for temperatures lower than $\sim 2\times10^4 K$.

   \begin{equation}
      \Lambda=n_e(n_p+n_H)\lambda(T)
   \end{equation}
where
   \begin{equation}
      \lambda(T)=10^{-21}10^{[log_{10}(T/T_M)]^2}\chi(T)
   \end{equation}
and
\[
\chi(T) =
\begin{cases}
1 & \text{for } T > T_1, \\
\frac{T - T_0}{T_1 - T_0} & \text{for } T_0 < T < T_1, \\
0 & \text{for } T < T_0.
\end{cases}
\]

$T_0$,$T_1$,$T_M$ are fitting parameters adjusted to simulate radiative losses and medium opacity. In the chromosphere, our formulation of radiative losses ensures that the temperature cannot exceed a threshold value defined here by $T_0=6500K$. $T_1$ and $T_M$ define the temperature range of the transition region. We choose $T_1=2\times10^4 K$ and $T_M=2\times10^5 K$.

\section{Constraining free parameters with Solar Orbiter data}
\label{sec:solo}

The first solar encounters of the Parker Solar Probe and Solar Orbiter missions have offered an exceptional context for testing ISAM against a multitude of datasets. We provide further tests here by exploiting the results from a recent magnetic connectivity study of Solar Orbiter data carried out by \cite{dakeyo_testing_2024}. In their work, proton measurements from the Proton-Alpha Sensor (PAS) \citep{owen_solar_2020,louarn_multiscale_2021} were used in combination with interplanetary magnetic field observations from the magnetometer MAG \citep{Horbury_MAG2020}, and a magnetostatic model of the corona and interplanetary magnetic field to connect wind properties to their coronal sources. \\

\begin{figure}[!h]
      \centering
      \adjustbox{margin=-0.375cm 0cm 0cm 0cm}{ 
      \includegraphics[scale=0.23]{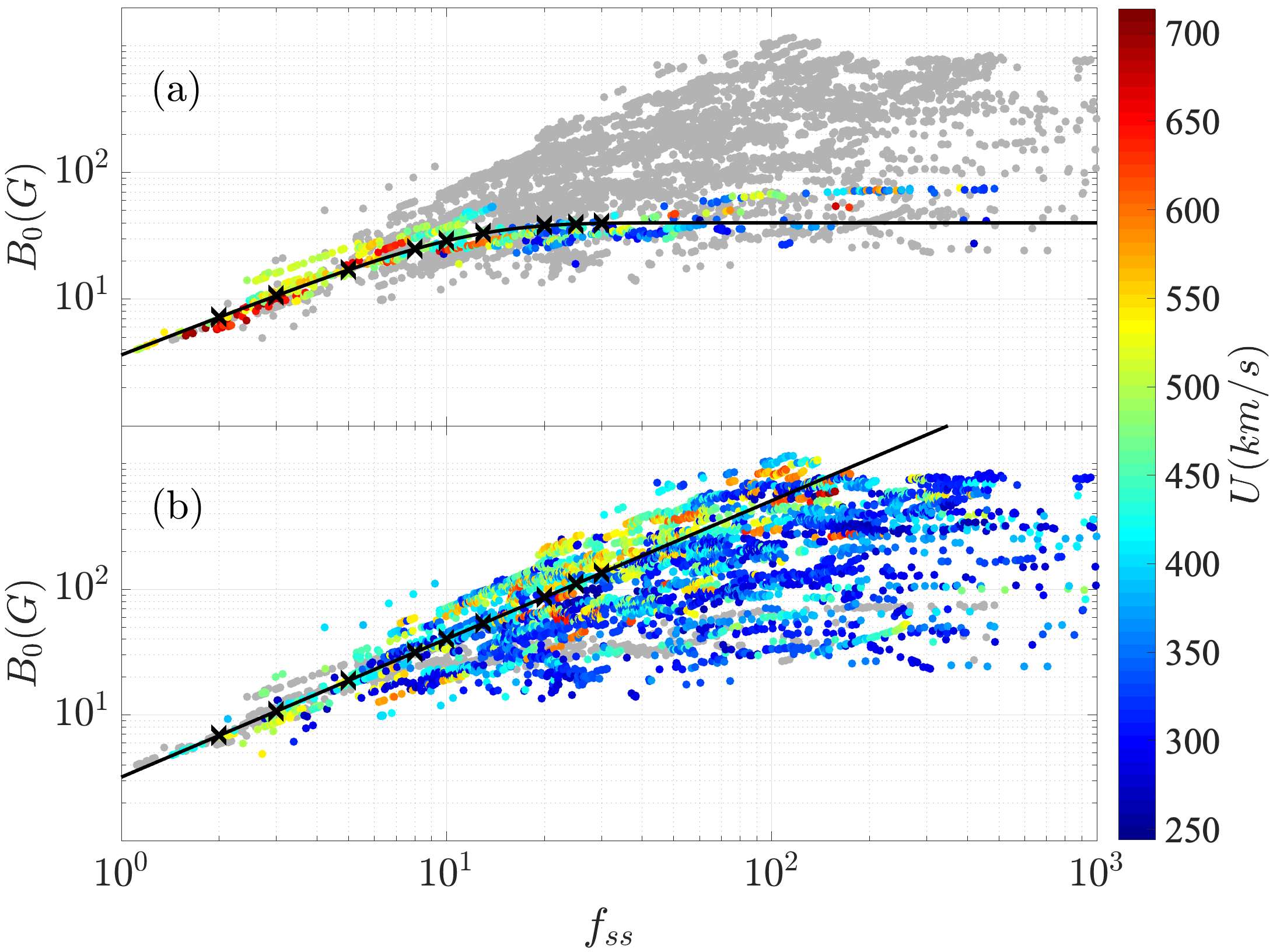}}
      \caption{Scatter plots of the photospheric magnetic field strength $B_0$ measured by Solar Orbiter, and the expansion factor $f_{ss}$ computed from a PFSS model \citep{dakeyo_testing_2024}. The data shown cover from 01/08/2020 to 17/03/2022 corresponding to the end of the solar minima. The values of the associated solar wind speeds at Solar Orbiter are displayed by the color code defined by the color bar. Panel a: $B_0-f_{ss}$ relations for connectivity points rooted at high latitudes ($>45^\circ$). The black line corresponds to the relation $B_0=f(f_{ss})$ used for our high latitude (HL) simulation series (Eq \ref{eq:Bhl}). Panel b: the $B_0-f_{ss}$ relations for connectivity points rooted at low latitudes ($<45^\circ$). The black line corresponds to the relation $B_0=f(f_{ss})$ used for our (LL) low latitude simulation series (Eq \ref{eq:Bll2}). For all panels, the black crosses correspond to the $B_0$ associated with our $f_{max}$ sample.}
      \label{fig:FsB0}
  \end{figure}

The free parameters of the model, related to the geometry of the solar magnetic field (Eq \ref{eq:AA0}) and the heating terms described in Section \ref{sec:num}, can be further constrained on the basis of this Solar Orbiter study. This is illustrated in Figure \ref{fig:FsB0} which presents the relation between the photospheric field strength ($B_0$) as a function of the expansion factor ($f_{max}$) of magnetic field lines connected to each solar wind parcel measured by PAS. The wind velocity ($U$) measured by PAS is also indicated by the color scale. These measurements cover from 01/08/2020 to 17/03/2022, which corresponds to the rising phase of the solar minima. \\

The expansion factor computed by \cite{dakeyo_testing_2024} and appearing in Figure \ref{fig:FsB0} is calculated with a Potential Field Source Surface (PFSS) extrapolation of solar magnetograms \citep{altschuler_magnetic_1969}. In PFSS models, the magnetic field is assumed radial at the photosphere and at a theoretical spherical surface of heliocentric radius $r_{ss}$ called the source surface. Typically, this radius is taken to be $r_{ss}=2.5R_\odot$ to mimic the radial rays of solar eclipses \citep{wang_why_1991}. Since the field remains radial beyond that surface, the expansion factor $f_{ss}$ of the magnetic field between the photosphere and the source surface is exactly equal to $f_{max}$. We will therefore use $f_{ss}$ for the rest of this study to indicate the asymptotic value of the expansion factor.\\

Figure \ref{fig:FsB0}a reveals a positive correlation between the magnetic footpoint strength $B_0$ and the expansion of the field line $f_{ss}$, especially for lower $f_{ss}$ for which the correlation is clearer with a smaller scattering. This dependence of $f_{ss}$ on  $B_0$ means that the total energy flux term defined as $F_{B_0}^{Tot}=F_{B_0}^{cor}+F_{B_0}^{TR}$ and the coronal heating scale height are not independent parameters. 
$B_0$ is less and less correlated to $f_{ss}$ as $f_{ss}$ increases and the Figure \ref{fig:FsB0}b reveals that solar winds sharing similar $f_{ss}$ can originate on magnetic flux tubes with very different $B_0$. \\

The correlation between the footpoint field strength and the expansion factor  shown in Figure 2 is expected  from equation \ref{eq:AA0} since $B(r_{ss})$ undergoes little variation in comparison with $B_0$. The reason $B_0(f_{ss})$ in Fig. \ref{fig:FsB0}a levels off much more sharply than in Fig. \ref{fig:FsB0}b is also expected since source regions located at high latitudes  are not associated with strong magnetic fields unlike the active region belts.\\


Most of the wind parcels originating from high-latitude (polar) coronal holes points, shown in Figure \ref{fig:FsB0}a, follow the well-known anti-correlation between wind velocity and expansion factor \citep{wang_solar_1990} as noted in \citet{dakeyo_testing_2024}. This anti-correlation is however not as strongly verified for wind parcels originating from low-latitude (Figure \ref{fig:FsB0}b). This is supported by past magnetic connectivity studies \citep{wang_2016}, which compared the coronal magnetic field with solar wind properties and found very little correlation between $B_0$ and the solar wind speed measured near Earth.\\

The upper part of the $B_0(f_{ss})$ distribution of Figure \ref{fig:FsB0}b is characterised by the presence of fast winds originating from low latitude active region belt. These solar wind streams could undergo a different acceleration regime.\citet{dakeyo_testing_2024} showed that these winds become supersonic inside the region of magnetic field super-radial expansion, well below the source surface. They relate these unusually high speeds as the result of an early supersonic flow which is further accelerated by a de Laval nozzle effect \citep{seifert_1947} in the strongly diverging field (very high $f_{ss})$ in the low corona. We will show that this early supersonic flow is driven by a combination of strong heating at the coronal base (due to high $B_0$) and significant flux tube expansion (de Laval nozzle effect).\\

PFSS reconstructions should be used with caution especially during (1) elevated activity levels, (2) in the direct vicinity of active regions or (3) near streamer structures and their associated current sheets. For this reason we consider statistical trends from Fig. 2 instead of individual points that may be prone to significant backmapping errors. An analysis of the uncertainties in backmapping was carried out in \citet{dakeyo_testing_2024} and we refer the reader to their appendix A for further information. During the preparation of their study they also compared the estimated source locations directly with ultraviolet imaging of corona holes (private communication) and confirmed that the wind population associated with moderate wind speeds of $400-600 km/s$ and elevated expansion factors are rooted in low-latitude narrow (less than 5 degrees in width) coronal holes with strong magnetic field strengths. This magnetic connectivity was found to be long-lasting (1-2 days) due to the spread of expanding field lines. One should note that low-latitude coronal holes produce typically slow solar wind especially during solar maximum periods.\\

The high $U$, high $f_{ss}$ values are related to special conditions at the source of the solar wind. The present paper investigates whether very strong base heating related to high surface magnetic fields (input energy flux) could be the source of extra acceleration. This was of course not necessarily expected since strong base heating induces higher mass flux evaporated into the tube. What we find is that for the strongest base heating (highest values of $B_0$) the enhanced mass flux in the tube contributes to saturate the maximum speed in the range of $500-550 km/s$ which corresponds roughly to the fastest winds measured by Parker Solar Probe and Solar Orbiter close to the Sun.\\

The correlations shown in Figure \ref{fig:FsB0} can be used to reduce the number of free parameters of our model by providing relations between base heating $B_0$ and field geometry $f_{ss}$. To this end, we fitted two different parts of the distribution to capture the variety of observed solar wind conditions associated with $B_0$-$f_{ss}$ relations. The two fits follow the lowest and highest part of the $B_0(f_{ss})$ distribution, corresponding respectively to a high latitude simulation series (named \textit{HL}) and low latitude simulation series (\textit{LL}).\\

The parametric equation describing these fit are :
\begin{equation}
    B^{HL}_0=70\times\tanh\left(\frac{f_{ss}}{20}\right) \,\,\, G
    \label{eq:Bhl}
\end{equation}
\begin{equation}
    B^{LL}_0=3.2\times f_{ss}^{1.1} \,\,\, G
    \label{eq:Bll2}
\end{equation}

These relations, combined with our heating prescription described in Section \ref{sec:heating_rates}, make $f_{ss}$ the free parameter that defines not only the field geometry and its effect on the divergence terms in the conservation equations, but also the heating scale height $H_f$ and the energy flux $F_{B_0}$ through $B_0$. \\

Each simulation series to follow in the next section is made of nine simulations corresponding to $B_0$-$f_{ss}$ pairs indicated by the black crosses in Figure \ref{fig:FsB0}. The nine simulations correspond to nine values of the expansion factors $f_{ss}={2,3,5,8,10,13,20,25,30}$. The three simulations series share the same three first $B_0$-$f_{max}$ values situated on the superposed portions of the black blue curves.\\

\section{Wind simulations for high-latitude coronal holes}
\label{sec:HL}

      \begin{figure*}[!h]
          \centering
          \includegraphics[scale=0.42]{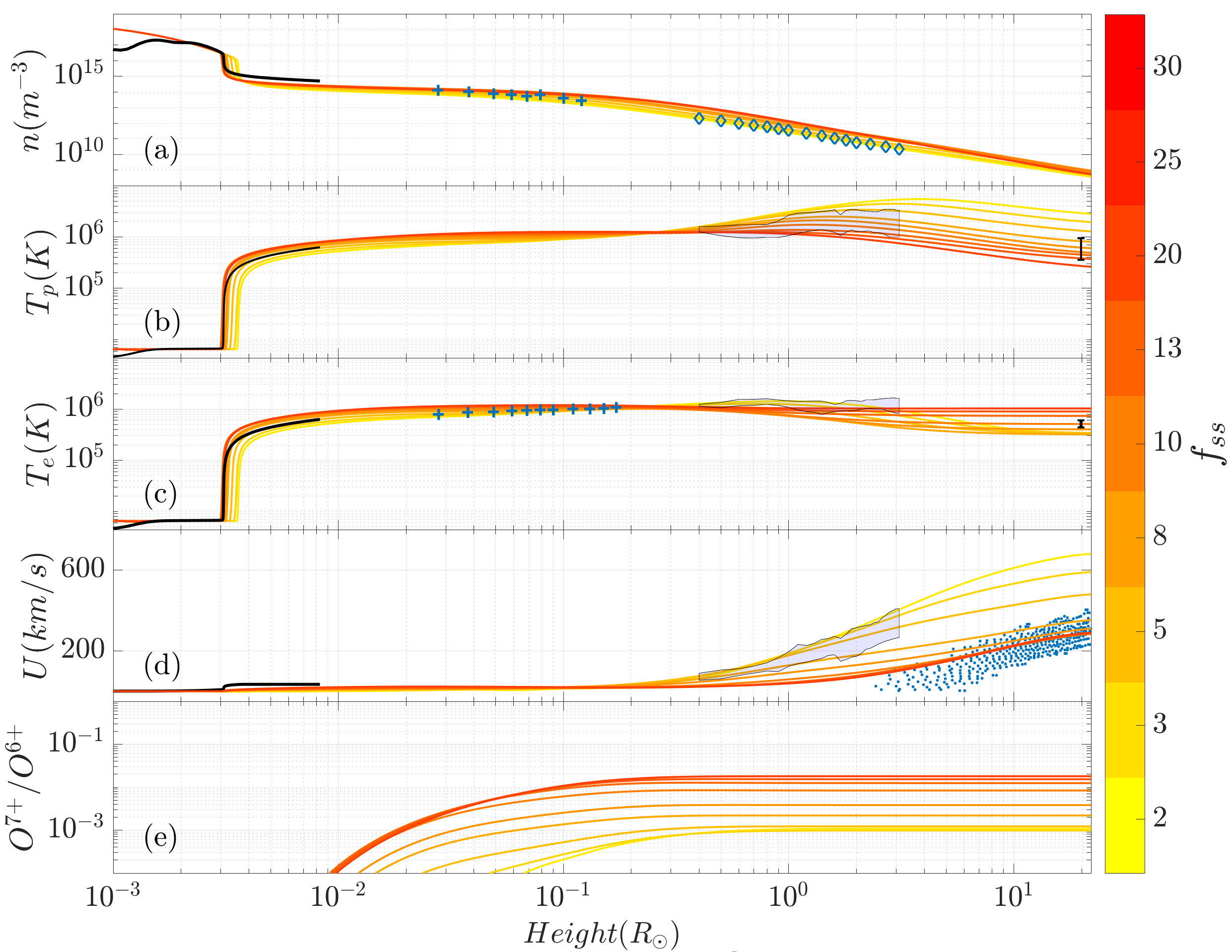}
          \caption{\emph{Simulation results for the high latitudes (HL) series as a function of height above the photosphere.\/} The colormap indicates the $f_{ss}$ value associated with each simulation. Panel (a) :  Proton density compared with observations : blue crosses correspond to SOHO-SUMER off-limb measurement from \cite{landi_off-disk_2008}, blue diamonds correspond to UVCS measurement from \cite{cranmer_heating_2020}. Panel (b) :  Proton temperature compared with observations : grey patches correspond to UVCS measurement from \cite{cranmer_heating_2020}, black vertical lines with caps corresponds to measurement of Parler Solar Probe \citep{dakeyo_statistical_2022}.  Panel (c) :  Electron temperature compared with observations : blue triangles correspond to SOHO-SUMER off-limb measurement \citep{landi_off-disk_2008}, grey patches correspond to UVCS measurement from \cite{cranmer_heating_2020}, black vertical lines with caps corresponds to measurement of Parler Solar Probe \citep{dakeyo_statistical_2022}. Panel (d) :  Proton velocity compared with observations : grey patches correspond to UVCS measurement from \cite{cranmer_heating_2020}, blue dots between correspond to blobs velocity in the HCS measured by SOHO/LASCO and by the Solar TErrestrial RElations Observatory STEREO/COR \citep{abbo_slow_2016}. Panel (e) :  Oxygen charge state ratio $n_{O^{7+}}/n_{O^{6+}}$ calculated using the method described in appendix \ref{ax:ko_ioniz}. For Panel (a)-(d) : black profiles in the chromosphere correspond to non-LTE radiative semi-empirical model from \citet{avrett_models_2008}. 
      }
      \label{fig:HLWindSimu}
      \end{figure*}
\begin{table*}[t]
\centering
\begin{tabular*}{\linewidth}{@{\extracolsep{\fill}} c|ccccccccc}
\hline
 \text{Input parameters}& & & & & & & & & \\ 
\hline
 $f_{max}$&2&3&5&8&10&13&20&25&30\\ 
 $B_0$ (G)&2.9&4.3&7&10.9&13.2&16.4&21.8&24.3&25.9\\ 
 $H_f^{TR}$ ($10^{-2}R_\odot$)&1&1&1&1&1&1&1&1&1\\ 
 $H_f^{cor}$ ($R_\odot$)&2.6&1.9&1.3&0.9&0.77&0.63&0.46&0.39&0.33\\ 
 $F_{B_0}^{TR}$ ($10^{5} \ erg \ cm^{-2} \ s^{-1}$)&1.25&1.85&3&4.69&5.7&7.1&9.4&10.5&11.1\\ 
 $F_{B_0}^{cor}$($10^{5} \ erg \ cm^{-2} \ s^{-1}$) &2.5&3.7&6&9.38&11.4&14.2&18.8&21&22.2\\ 
 $r_p^{sw}$ ($R_\odot$)&4.80&3.6&2.5&1.68&1.5&1.2&0.8&0.7&0.6\\ 
 $r_e^{sw}$ ($R_\odot$)&6.8&4.3&2.6&1.63&1.3&1.0&0.6&0.5&0.4\\ 
\hline
\end{tabular*}
\caption{Input parameters for the different simulations within the high latitude series (HL) calculated using 
Eq \ref{eq:Fcor}, \ref{eq:Hcor},\ref{eq:Ftr}, \ref{eq:Bhl}.  $r_e^{sw}$ and $r_p^{sw}$ have been chosen to fit PSP data at $20R_\odot$}
\label{table:InputparamHL}
\end{table*}


We begin by modeling the solar wind emerging from a polar coronal hole by considering the HL simulation series defined in the previous section. The lower boundary conditions and all input parameters are shown in Table \ref{table:InputparamHL}. The simulation results are compared to remote sensing and in-situ observations in Figure \ref{fig:HLWindSimu}. \\ 

The profiles extend from the highly collisional photosphere at $1\times10^{-3}R_\odot$ ($\sim 700$ km) to $20R_\odot$, in the collisionless solar wind. The transition region visible as the sharp transition in plasma moments located near $3\times10^{-3}R_\odot$($\sim 2000$ km) agrees closely with the temperature and density jumps given by the non-Local Thermodynamic Equilibrium (LTE) radiative model of \citep{avrett_models_2008} shown as black profiles in the different panels.\\

Increasing $f_{ss}$ reduces $H_f$ which has for main effect to concentrate coronal heating below the sonic point, in the still collisional part of the corona. Heating in this region increases the base temperature, the conductive heat flux towards the transition region, thereby raising its pressure and chromospheric evaporation. This increases mass flux and therefore the density in the low corona. These effects are visible in panels (a)-(d) where $n$, $T_{p}$, $T_{e}$ and $U$ all increase in the lower corona with increasing values of $f_{ss}$. The increased coronal pressure pushes the transition region Sunward and the best correspondence between our model and the \cite{avrett_models_2008} profiles is obtained for the highest values of $f_{ss}$.\\

Concentrating energy deposition in the lower atmosphere decreases the energy available for solar wind acceleration in the upper corona. The highest terminal wind speeds are obtained for the smallest expansion factor values $f_{ss}={2,3,5}$, the associated curves agree with the temperatures and speeds derived from spectroscopic observations made with SoHO/UVCS of polar coronal holes \citep{cranmer_heating_2020}. \\    

Increasing $f_{ss}$ results in a progressive decrease in wind speed and pushes the bulk of the wind acceleration higher up in the solar atmosphere. The lowest terminal wind speeds are for $f_{ss}={20,25,30}$ and their acceleration follows closely that of transient structures released in the heliospheric current sheet observed as 'blobs' by coronographs onboard SoHO and STEREO \citep{abbo_slow_2016} (shown as blue dots in Figure \ref{fig:HLWindSimu}d). We do not expect wind velocities to be lower than that of blobs considered 'passive' structures entrained in the slow wind \citep{sheeley_1997, reville_tearing_2020}.\\

Collisions in the lower corona maintain $T_e \approx T_p$ until the medium becomes collisionless and $T_e$ and $T_p$ begin to differ near $3\times 10^{-1}R_\odot$ ($\sim 2000$ km) for the fastest winds but at higher altitudes for the slower winds. The simulations shown in Figure \ref{fig:HLWindSimu} also retrieve the well-known correlation between $T_p$ and $U$ and the anti-correlation between $T_e$ and $U$ in the escaping solar wind \citep{dakeyo_statistical_2022}. Comparing $T_e$ and $T_p$ at the outer boundary with particle measurements by PSP (black vertical lines with caps shown in panels b and c) provided by \citep{dakeyo_statistical_2022}, we see that the simulations slightly overestimate the range of temperatures.\\

We also note from Figure \ref{fig:HLWindSimu}b,d that for the fast wind to have speeds of at least $\sim 600$ km/s, simulated proton temperatures must exceed those inferred in coronal holes from spectroscopy as well as in the nascent solar wind measured in-situ by PSP. This points to the well-known result that thermal pressure gradients are not sufficient to produce the highest wind speeds measured in-situ and additional sources of momentum must be considered \citep{pierrard_acc_2024,cranmer_self-consistent_2007}. This missing momentum source will be addressed in section \ref{sec:solu_supra}.\\

\section{Source temperatures at high latitudes}
\label{sec:cs_hl}

\begin{figure}[!h]
      \centering
       \adjustbox{margin=-0.4cm -1cm 0cm 0cm}{
      \includegraphics[scale=0.265]{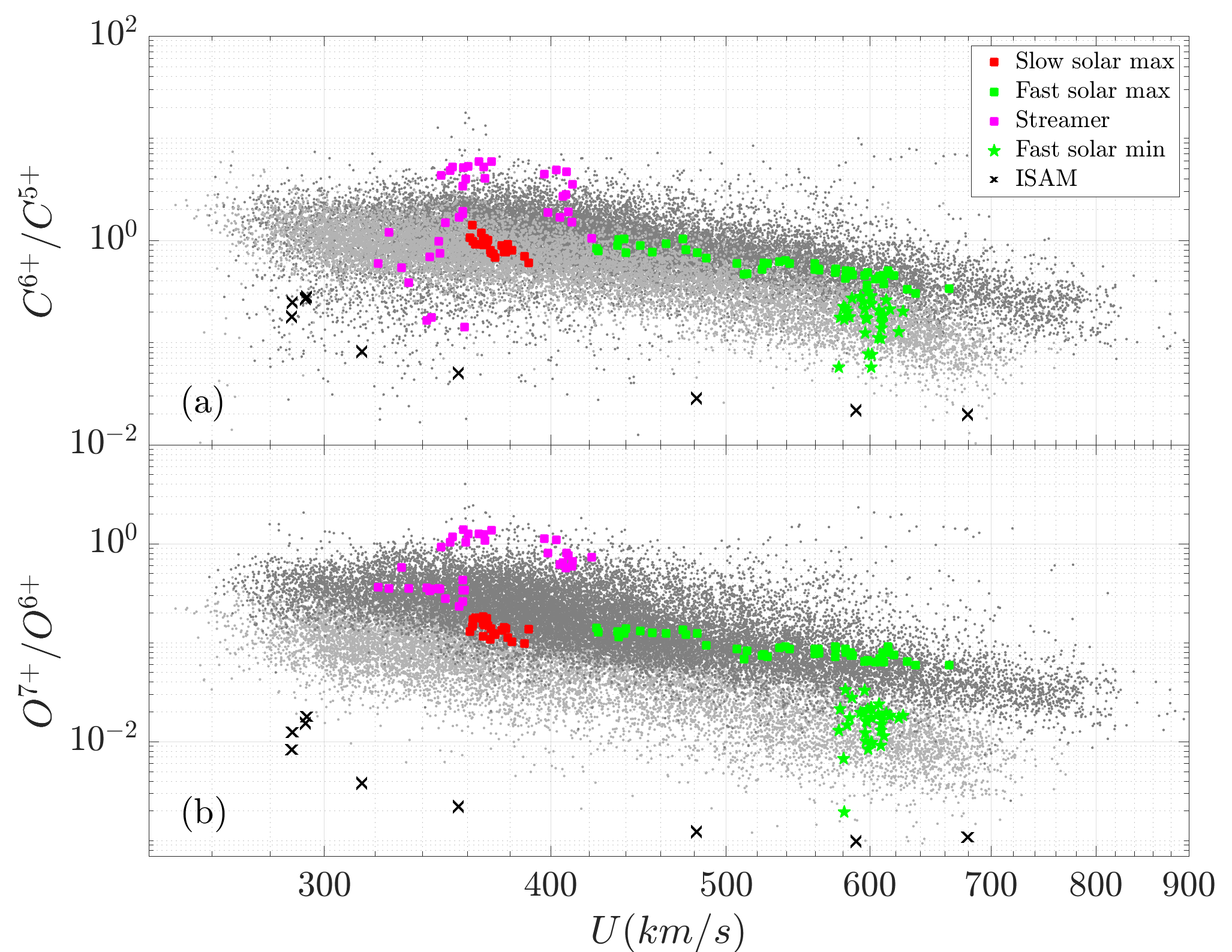}}
          \caption{\emph{\ In-situ measurements by ACE of the ion CSRs $n_{O^{7+}}/n_{O^{6+}}$ (panel a) and $n_{C^{6+}}/n_{C^{5+}}$ (panel b) as a function of the proton wind speed compared with solutions of equation \ref{eq:d_y}.} 
          The light grey dots represent the solar minimum corresponding to the period 04/09/2006 to 24/09/2009. The dark grey dots represent the solar maximum corresponding to the period 07/04/1999 to 22/12/2003. Black crosses correspond to solutions of equation A.1 associated to the high latitude (HL) simulation series. The data corresponding to the time intervals of the fast wind, Alfvènic slow wind, and non Alfvènic slow wind identified in the study of \cite{damicis_origin_2015} are shown as green, red and black dots and stars respectively. 
      }
      \label{fig:CSRobsmodHL}
  \end{figure}

Figure \ref{fig:CSRobsmodHL} presents in situ measurements by the Advanced Composition Explorer (ACE) of the oxygen charge state ratios $n_{O^{7+}}/n_{O^{6+}}$ for the period 06/02/1998 to 21/08/2011. The distribution exhibits the well-known anti-correlation between source temperature ($n_{O^{7+}}/n_{O^{6+}}$) and wind speed introduced earlier which implies a seemingly hotter source for slower solar winds. The slow wind also tends to have a broader range of possible source temperatures if we consider the plot is in logarithmic scale. As already discussed, this is consistent with the multi-component nature of the slow wind indicating different possible conditions at the source and/or different formation mechanisms. The highest charge state ratios (CSRs) come from slow winds originating in the vicinity of coronal streamers where enhanced charge states could result from coronal loops material ejected into the solar wind by reconnection with open field lines which may provide an explanation for their high value \citep{rouillard_solarwind_2021,sheeley_1997,sanchezdiaz_2017,reville_tearing_2020, reville_flux_2022}. In the framework of the quasi-static theory, the modeling of these winds is not considered. \\

The periods identified in \cite{damicis_2015} of slow winds sharing similar Alfvénicity and compressibility to the fast wind are plotted as red dots in Figure \ref{fig:CSRobsmodHL}. The fast wind periods also identified by these authors during this solar maximum period are shown as green dots. These fast and slow winds could be considered a continuum determined by the expansion rate of the coronal magnetic field. For these two wind components, we can still observe an anti-correlation of the CSRs with wind speed that can be investigated using our HL ISAM simulations tested in the previous section against a broad range of observations.\\

Charge-state ratios $n_{O^{7+}}/n_{O^{6+}}$ were calculated from the ISAM simulations in two different ways in the present paper: (1) by using the electron moments modelled by ISAM as input to a solution of the system of continuity equations in \citet{ko_empirical_1997} (detailed in Appendices \ref{ax:ko_ioniz} and \ref{ax:kappa_ioniz}), (2) by solving for the full transport and ionisation of heavy ions treated as an additional population in ISAM. The two approaches gave similar results that are presented in Appendix \ref{ax:oxygen}. A direct solution of the heavy ion transport accounts for differential flows that are found to induce slightly higher ionisation levels than without as discussed in Appendix \ref{ax:oxygen}.\\

The charge-state ratios shown in Figure \ref{fig:HLWindSimu}(e) and also reported in  Figure \ref{fig:CSRobsmodHL} were therefore computed using the simpler and less computationally intensive method of \citet{ko_empirical_1997}. As seen in previous studies, the $n_{O^{7+}}/n_{O^{6+}}$ evolves rapidly near the transition region and settles below $1R_\odot$ remaining invariant further out. Increasing $B_0$ and its associated input energy flux, $n_{O^{7+}}/n_{O^{6+}}$ increases in response to the induced rise in $n$ and $T_e$. The wind velocity $U$ increases between consecutive simulations but it is not sufficient to decrease $n_{O^{7+}}/n_{O^{6+}}$ via non-equilibrium effects. \\

These simulations illustrate how a displacement of energy deposition towards the base of the corona decreases the terminal wind speed, increases the coronal base temperature and density thereby increasing the charge state of heavy ions. This was already shown by \citep{wang_slow_2009} and is verified here further through a more comprehensive and systematic comparison with a broad range of datasets. We note from Figure \ref{fig:CSRobsmodHL} that the simulated CSRs match the lower values of the distributions but generally lower than those measured in situ. This discrepancy will be revisited in section \ref{sec:sim_ar} and \ref{sec:solu_supra}.

      \begin{figure*}[!h]
          \centering
          \includegraphics[scale=0.43]{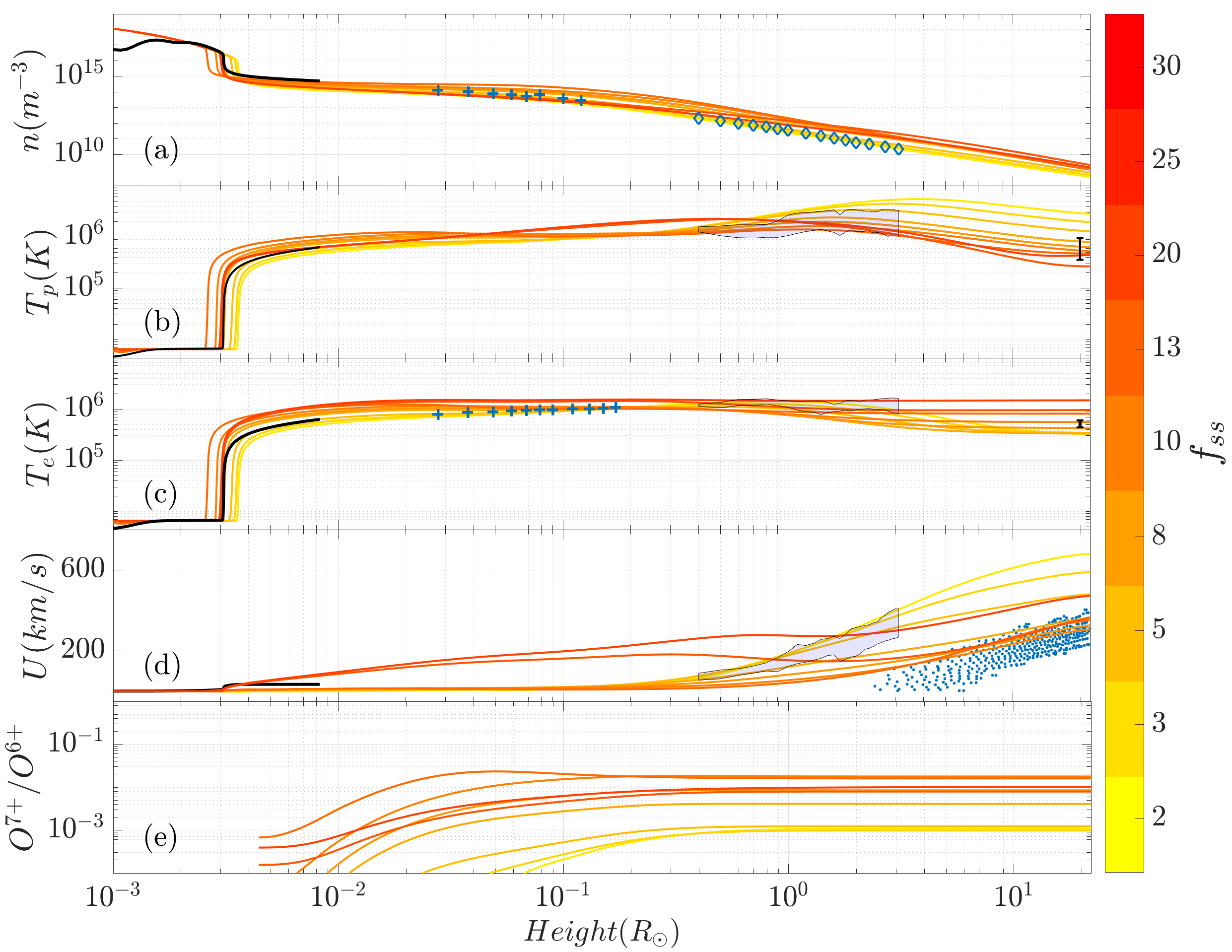}
          \caption{\emph{Simulation results for wind sources rooted at low latitudes (LL) as a function of height above the photosphere.\/} Comparison with observations and models are the same as in Figure \ref{fig:HLWindSimu}.
      }
      \label{fig:LL2}
      \end{figure*}

\section{Simulating wind sources from the active region belt}
\label{sec:sim_ar}
A significantly enhanced input energy flux is expected for wind sources located within the active region belt where the strongest photospheric magnetic fields are encountered. Such conditions correspond to the $B_0(f_{ss})$ values along the upper curve (LL) shown in Figure \ref{fig:FsB0}b. Along this branch, $B_0$ values at $f_{max}=25,30$ are over an order of magnitude higher than for the HL runs as reported in Table \ref{table:InputparamLL2}. \\

Increasing  $B_0$, and therefore the related input energy flux, forces the wind to transition into a new expansion regime. For the subset of simulations related to the highest values of $B_0(f_{ss})$, the wind speeds in the low corona are much higher than for similar $f_{ss}$ values in the HL simulations (Figure \ref{fig:HLWindSimu}). Supersonic speeds develop already in the region of strong super-radial field expansion in the low corona. \cite{kopp_dynamics_1976} and \cite{dakeyo_testing_2024} studied the effect of such flux-tube geometry on the wind velocity profiles using polytropic models and a combination of isothermal and polytropic modeling, respectively. They showed that when the sonic point is brought below or within the region of super-radial expansion, the solar wind experiences a significant acceleration. \\

For such a high regime of $B_0$ and $f_{ss}$ most of the energy is deposited in the low corona and the wind accelerates only gradually in the upper corona to reach moderately high speeds. Nevertheless, this regime explains the occurrence of the faster wind speeds ($>450$ km/s) observed for high $f_{max}$ in Figure \ref{fig:FsB0} and discussed in detail in \cite{dakeyo_testing_2024}.\\ 

The strong base heating induces some significant chromospheric evaporation into the corona. However, the increased plasma acceleration and the high $f_{ss}$ induce a significant expansion of the plasma acting to decrease the density and limiting the rise in temperature in the low corona. Low densities and increased temperatures lead to a drop in the collision frequencies allowing a strong divergence of electron and ion temperatures. The electrons rapidly decouple from the ions which reduces the energy transmitted by collisions to the ions, resulting in a decrease in the ionic temperature between $3\times10^{-3}$ ($\sim 2000$ km) and $4\times10^{-2}$ ($\sim 27 000$ km). \\

According to equation \ref{eq:d_y}, the increase in wind speed and the decrease in density limit the rise in the ionization level. This can be seen in the profiles corresponding to the second expansion regime ($f_{ss}=25,30$) in Figure \ref{fig:LL2}) where the high velocities (up to $\sim 200 km/s$) and low densities in the collisional region of the corona result in oxygene CSRs no longer following the increasing trend, falling below the charge ratios associated with the $f_{ss}=13,20$ profiles. \\

\begin{figure}[!h]
      \centering
       \adjustbox{margin=-0.4cm 0cm 0cm 0cm}{
      \includegraphics[scale=0.25]{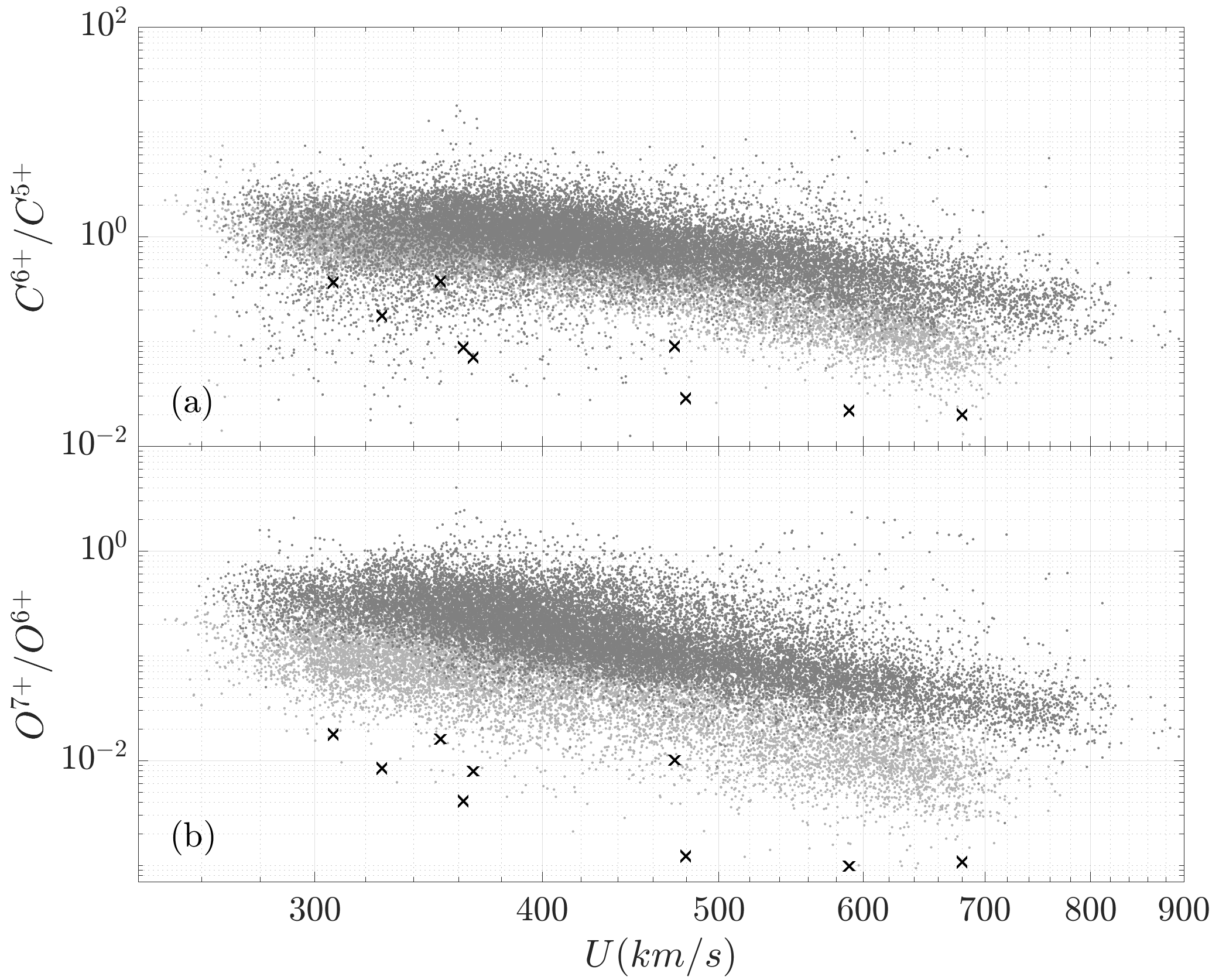}}
          \caption{\emph{\ In-situ measurements by ACE of the ion CSRs $n_{O^{7+}}/n_{O^{6+}}$ (panel a) and $n_{C^{6+}}/n_{C^{5+}}$ (panel b) as a function of the proton wind speed compared with solutions of equation \ref{eq:d_y}.} 
          The light gray dots represent the solar minimum corresponding to the period 04/09/2006 to 24/09/2009. The dark gray dots represent the solar maximum corresponding to the period 07/04/1999 to 22/12/2003. Black crosses correspond to solutions of equation A.1 associated to the low latitude (LL) simulation series.
          }
      \label{fig:cs_sans_k}
  \end{figure}

The carbon and oxygen CSRs from our LL simulation series were calculated from equation \ref{eq:d_y} (as for the HL simulations series) and are compared with the in-situ ACE measurements in Figure \ref{fig:cs_sans_k}. 

The two $B_0(f_{ss})$ relations (Figure \ref{fig:FsB0}) considered for the high-latitude (HL) (Figure \ref{fig:HLWindSimu}) and low-latitude (LL) simulation (Figure \ref{fig:LL2}) provide the lower and upper limits of base heating conditions and of possible CSRs. These simulations provide an interpretation for the anti-correlation between wind velocity and CSRs. However, as with other attempts to model CSRs with thermal models \citep{cranmer_self-consistent_2007,oran_steady-state_2015}, CSRs values are below those measured in situ. Even the LL series winds with the highest electron temperatures at the base provide CSR values that are below the center of the distribution. \\

Past modeling studies obtained even lower $n_{O^{7+}}/n_{O^{6+}}$ values below observations by a factor 10, demonstrating further that thermal models are unable to retrieve the observed charge states \citep{cranmer_self-consistent_2007,cranmer_suprathermal_2014,oran_steady-state_2015}. Additional processes should be considered and we now consider the combined effects of non-thermal particles on the solar wind properties and charge-state ratios.
\\

\begin{table*}[t]
\centering
\begin{tabular*}{\linewidth}{@{\extracolsep{\fill}} c|ccccccccc}
\hline
 \text{Input parameters}& & & & & & & & & \\ 
\hline
 $f_{max}$&2&3&5&8&10&13&20&25&30\\ 
 $B_0$ (G)&2.9&4.3&7&13.8&17.6&23.5&37.8&48.3&59.0\\ 
 $H_f^{TR}$ ($10^{-2}R_\odot$)&1&1&1&1&1&1&1&1&1\\ 
 $H_f^{cor}$ ($R_\odot$)&2.6&1.9&1.3&0.9&0.77&0.63&0.46&0.39&0.33\\ 
 $F_{B_0}^{TR}$($10^{5} \ erg \ cm^{-2} \ s^{-1}$)&1.25&1.85&3&5.54&7.6&10.1&16.25&20.52&25.35\\ 
 $F_{B_0}^{cor}$($10^{5} \ erg \ cm^{-2} \ s^{-1}$)&2.5&3.7&6&11.8&15.2&20.2&32.5&41.5&50.7\\ 
 $r_p^{sw}$ ($R_\odot$)&4.80&3.6&2.5&1.68&1.5&1.2&0.8&0.7&0.6\\ 
 $r_e^{sw}$ ($R_\odot$)&6.8&4.3&2.6&1.63&1.3&1.0&0.6&0.5&0.4\\ 
\hline
\end{tabular*}
\caption{Input parameters for the different simulations within the low latitudes series (LL) calculated using Eq \ref{eq:Fcor}, \ref{eq:Hcor},\ref{eq:Ftr}, \ref{eq:Bll2}.}
\label{table:InputparamLL2}
\end{table*}

\section{Wind solutions including suprathermal electrons:}
\label{sec:solu_supra}

Non-thermal particle populations are ubiquitous in the solar wind as detected in numerous measurements of particle velocity distribution functions \citep{montgomery_solar_1968,feldman_solar_1975,pilipp_variations_1987,maksimovic_ulysses_1997}. Suprathermal electrons tend to follow a kappa or generalized Lorentzian velocity distribution in the solar wind \citep{vasyliunas_1968}. The high-energy tail then follows a power law in particle velocity:

\begin{equation}
    f_\kappa(v_e,\kappa)=n_e\left(\frac{m_e}{2\pi k_B T_e}\right)^{\frac{3}{2}}\frac{\Gamma(\kappa+1)}{\kappa^{\frac{3}{2}}\Gamma(\kappa-0.5)}\left[1+\frac{v_e^2}{\kappa\theta_e^2}\right]
    \label{eq:fk}
\end{equation}\\

where $v_e$ is the microscopic velocity of electrons, $\Gamma$ is the Gamma function, $\theta_e=\sqrt{\frac{2k_BT_e}{m_e}}$ is the thermal velocity of electrons and $\kappa$ is the parameter characterizing the amplitude of the supra-thermal population with $f_\kappa(v_e,\kappa=\infty)=f_{Maxwellian}(v_e)$. It is important that $\kappa > 3/2$ so that the integral of the distribution function converges and the moments are well defined.\\

Past modeling studies suggest the existence of suprathermal electrons already in the corona based on considerations of coronal turbulence \citep{roberts_generation_1998}, heating processes involving obliquely propagating finite-amplitude electromagnetic waves \citep{vinas_generation_2000},
nanoflares \citep{che_origin_2014} and resonant interactions with whistler waves \citep{vocks_generation_2003}. \\

Suprathermal electrons become collisionless in the mid-corona ($\sim 2\times 10^{-1} R_\odot)$, a subset of electrons have sufficient energy to overcome the Sunward directed electrical and gravitational forces and escape into the solar wind. These escaping electrons contribute to the acceleration of the solar wind via the ambipolar electric field \citep{pierrard_acc_2024}. Our thermal model based on bi-Maxwellian particle distributions could not produce terminal wind speeds of $\sim 700km/s$ from the measured coronal temperatures (see sections \ref{sec:HL} and \ref{sec:cs_hl}). Additional effects must occur and the ambipolar electric field generated by suprathermal electrons could be the missing force to accelerate the solar wind. In the low corona, high-energy electrons should also increase the CSRs of heavy ions \citep{cranmer_suprathermal_2014,oran_steady-state_2015,ko_limitations_1996}.
This section investigates jointly the effects of non-thermal electrons on the solar wind energy budget and ionisation levels of heavy ions.\\

Exospheric models of the forming solar wind are the simplest approach to model the evolution and effects of the electron distribution function. They assume the absence of Coulomb interactions between particles after a certain altitude, referred as the exobase \citep{Pierrard2021}. Each electron is therefore uniquely subject to external forces, including the electric, gravitational and Lorentz force. These simplifications combined with a stationarity condition and a predefined distribution function at the exobase permit an analytical solution of the Vlasov equation to get the radial profile of the velocity distribution function. Inclusion of the ambipolar electric force in ISAM was carried out by using the exospheric model of \cite{pierrard_exospheric_2023}, the modelled ambipolar electric field was included as an extra force in the momentum equation (\ref{eq:U}).\\

We included the ambipolar electric field to three characteristic simulations taken from our two sets of simulations: a fast ($f_{ss}=5$) and slow ($f_{ss}=20$) wind simulation taken from the HL set (representative of sources inside polar coronal hole), and a slow wind  ($f_{ss}=25$) simulation taken from the LL set (representative of sources located near active regions). We will refer to them respectively as $HL_{5},HL_{20},LL_{25}$. We used these three simulations to first compute the location of the exobase identified as the inflection point in the wind acceleration profile and found $\sim0.7R_\odot, 2R_\odot$ and $2.3R_\odot$ respectively for $HL_{5},HL_{20}$ and $LL_{25}$.  The ISAM profiles for $n$, $U$, $T_p$, $T_e$ taken at the exobase were then given as input for the exospheric model. The latter was run for different values of $\kappa$ providing radial profiles of the ambipolar electric field. To avoid the discontinuity caused by the abrupt inclusion of the electric field at the exobase in ISAM, we added a smoothly decreasing function to the electric field in the sunward direction.\\



\begin{figure}[!h]
      \centering
       \adjustbox{margin=-0.4cm 0cm 0cm 0cm}{
      \includegraphics[scale=0.293]{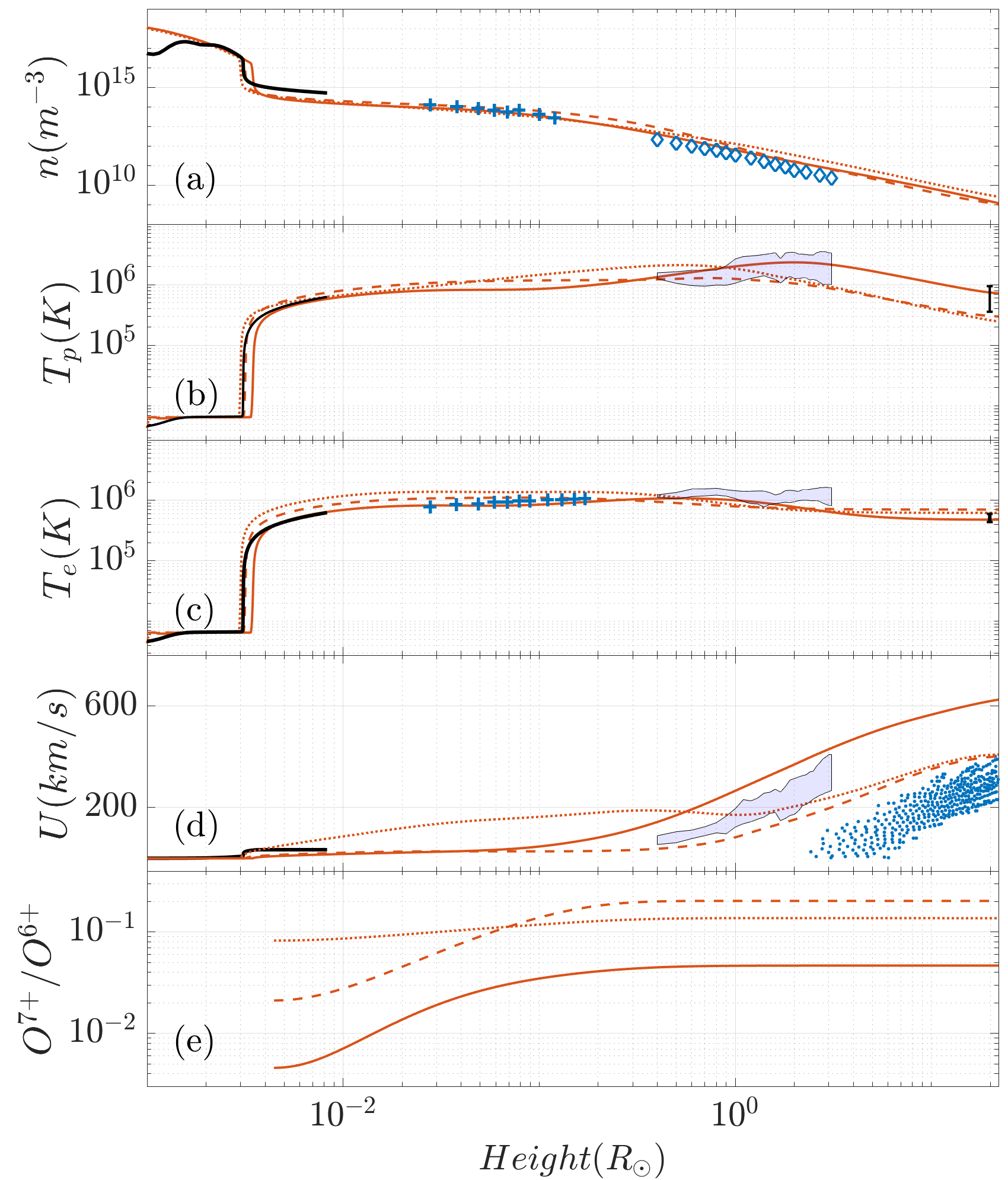}}
         \caption{\emph{\ Result of the simulation including the ambipolar electric field obtained with the exospheric model \citep{pierrard_exospheric_2023} and applied in ISAM for $\kappa=6$.} Solid, dashed and dotted red lines corresponds respectively to profiles from $HL_{5},HL_{20}$ and $LL_{25}$ simulations. The rest of the notation is the same as in Figure \ref{fig:HLWindSimu}.}
      \label{fig:simEfield}
     \end{figure}

A $\kappa=6$ provided the best fit between modeled and observed plasma moments as shown in Figure \ref{fig:simEfield}. As expected the additional momentum source provided by the applied ambipolar electric field induces wind speeds in excess of $\sim600 km/s$ for observed proton temperatures of $\sim 2 \times 10^{6} K$. A  $\kappa=6$ is consistent with the values $2<\kappa<6$ inferred by \citep{gloeckler_solar_1992, maksimovic_ulysses_1997} from electrons measured in-situ. 

We then calculate the CSRs for these 3 winds by taking into account the enhancement of ionisation associated with a kappa distribution with $\kappa=6$. To this end we follow an approach similar to \cite{ko_limitations_1996}. To use the ionisation/recombination rates pre-calculated for Maxwellian electron distributions, the kappa distribution is first approximated by a sum of Maxwellians, each characterised by a different density and temperature. The resulting ionisation and recombination rates are then the sum of the ionisation and recombination rates determined from the Maxwellians. The method is further described in Appendix
\ref{ax:kappa_ioniz}. \\

\begin{figure}[!h]
      \centering
      \adjustbox{margin=-0.4cm 0cm 0cm 0cm}{
      \includegraphics[scale=0.305]{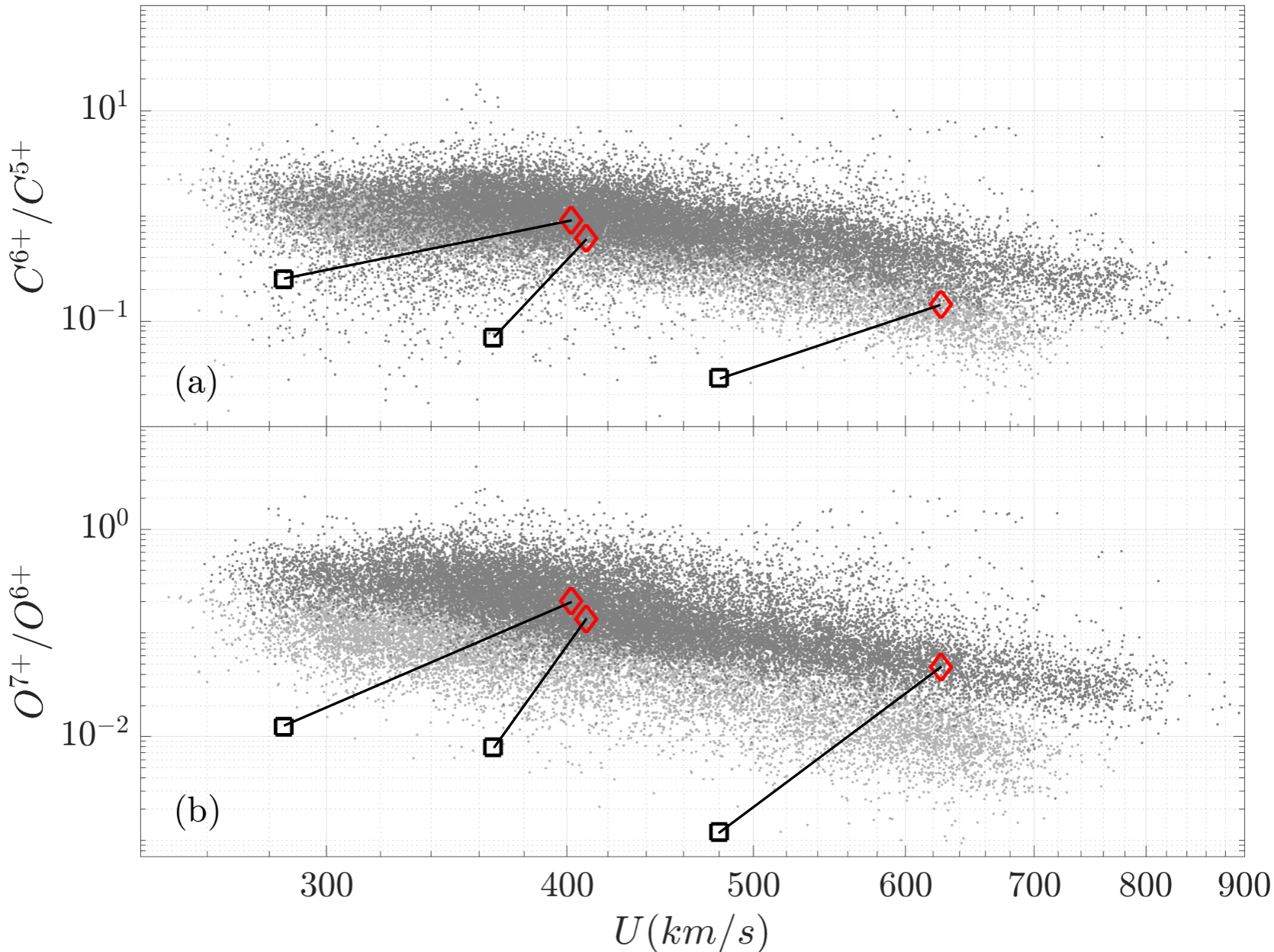}}
          \caption{\emph{\ Comparison of in-situ measurements by ACE of the ion CSRs at solar minima with solutions of equation \ref{eq:d_y}
          }. The light grey dots represent the solar minimum corresponding to the period 04/09/2006 to 24/09/2009. The dark grey dots represent the solar maximum corresponding to the period 07/04/1999 to 22/12/2003. Black squares correspond to solutions of equation \ref{eq:d_y} from $HL_{5},HL_{20}$ and $LL_{25}$ simulations assuming a Maxwellian distribution function. Red diamonds correspond  to solutions of equation \ref{eq:d_y} assuming kappa distribution function. The black lines linking the squares and diamonds link the simulations including suprathermal effects to their associated simulation without suprathermal effetcs. Panel (a) :  $n_{C^{6+}}/n_{C^{5+}}$ charge state ratio distribution
          Panel (b) : $n_{O^{7+}}/n_{O^{6+}}$ charge state ratio distribution
          }
      \label{fig:cs_avec_k}
  \end{figure}

The new CSRs values for our $HL_{5},HL_{20}$ and $LL_{25}$ simulations are reported in Figure \ref{fig:cs_avec_k} and compared with CSRs calculated using a Maxwellian eVDF from their associated simulations without the inclusion of ambipolar electric field. \\  

The addition of a suprathermal population has two effects visible in Figure \ref{fig:cs_avec_k} : the ambipolar electric field increases the terminal velocity of the winds and the high energy tail of the distribution enhances the ionisation and thus the CSRs. For the three simulations considered $HL_{5},HL_{20}$ and $LL_{25}$, we obtain a enhancement of CSRs  by a factor $\sim 3-10$ for ${C^{6+}}/{C^{5+}}$ and a factor $\sim 10-30$ for ${O^{7+}}/{O^{6+}}$. This results in a better fit of the oxygen CSR with solar maximum measurements, while the carbon CSR fits better with solar minimum measurements. Adding the effect of suprathermals makes it possible to recover CSR values that are consistent with the measurements, while retaining the characteristic anti-correlation between wind speed and CSRs.



\section{Discussion and Outlook}

In this study, a plasma transport model extending from the chromosphere to the solar wind was used to model the thermodynamic coupling between the different layers of the solar atmosphere, in the estimated source region of the solar wind measured by Solar Orbiter. Neutral particles were included to model ionisation processes and equilibria near the transition region. \\

The spatial distribution of coronal and solar wind heating was based on heating functions \citep{withbroe_temperature_1988, pinto_multiple_2017} that have been shown to match approximately the energy deposition by a turbulent cascade of Alfvén waves. In this respect, the spatial distribution of the energy deposition was made dependent on the large-scale magnetic field properties. The expansion rate of the magnetic field lines, along which the solar wind expands, controls the gradient in Alfvén speed, which in turn controls the reflection of Alfvén waves and the distribution of dissipation with height. The footpoint field strength was taken to be proportional to the Alfvén wave energy flux injected at the base of the tube. This corresponds to perpendicular velocity perturbations at the inner boundary. The expansion factor and the surface field strength, taken from the connectivity study of \citet{dakeyo_testing_2024}, were found to be correlated over ranges of low to medium expansion factor values ($f_{ss} \lesssim 20$). \\

With these constraints, our first goal was to verify that the model can reproduce recent in-situ measurements of solar wind properties by PSP and SolO in conjunction with a wide range of coronal source properties given by remote sensing observations. A parametric study was carried out, and we present a set of simulations that well reproduced wind properties originating from high and low latitudes. This work confirmed previous results showing that in models coupling the different layers of the low solar atmosphere, the height of the energy deposition strongly influences the wind properties through a number of mechanisms such as chromospheric evaporation, and thermal expansion above and below the sonic point. This work retrieves the well-known anti-correlation between wind speed and expansion factors, especially for solar wind originating from high latitudes. For the solar wind originating from low latitudes, i.e. footpoints in the vicinity of strong photospheric magnetic field, which is also associated to large coronal heating, the solar wind undergoes a significant expansion pushing the sonic point below the region of strong magnetic field expansion. We see that this regime leads to a strong acceleration of the solar wind in the low corona as already discussed by \citet{dakeyo_testing_2024} using their iso-poly wind model. This expansion regime could also explain the elevated plasma speeds ($\sim$ $100-200$ km/s) inferred near wind sources from remote-sensing observations \citep{Bemporad_2017} or the persistent outflows observed by coronal spectroscopy on the edge of active regions with velocities exceeding $100$ km/s as inferred by \citep{harra_outflow_2008}. Although the latter outflows have been interpreted as source locations of the solar wind and our simulations provide some theoretical support, it should be noted that they could also result from plasma transport along large-scale magnetic loops.\\

The aim of our study was to gain more insight into the mechanisms that control the charge state of heavy ions, or equivalently, the electron temperature, at the sources of the different types of solar winds. For the high latitude source regions, when energy deposition is concentrated at lower/upper heights in response to strong/weak magnetic flux tube expansions, the charge state increases/decreases. This naturally produces an anticorrelation between ion charge state and solar wind speed and remains clear if the surface field strength is not too strong. This interpretation was already proposed in \citet{wang_slow_2009} and explained in terms of Alfvén wave dissipation by \citet{cranmer_heating_2020}. We show in addition that the electron temperature at the low-latitude sources increases significantly due to the significant heating that occurs in these regions. However, the increase in charge state is limited by the significant plasma expansion. \\

Overall, the simulated ion charge states are slightly lower than those measured in the solar wind (Figure \ref{fig:cs_sans_k} and \ref{fig:CSRobsmodHL}). Since the coronal source temperatures can also be inferred from spectroscopic observations and are in agreement with the simulated ones (Figures \ref{fig:HLWindSimu} \& \ref{fig:LL2}), additional effects not included in our model must lead to higher charge states in the low corona. Our simulations also show that, at the observed coronal temperatures, a purely thermally driven solar wind cannot reach speeds above 600 km/s, in contrast to observations of the fast solar wind (Figure \ref{fig:HLWindSimu}). Both drawbacks could be overcome by inferring the presence of suprathermal electrons in the corona. Using an exospheric model, we show that these non-thermal particles help to increase the ionisation of heavy ions through collisions, as well as the acceleration of thermal protons through the development of a polarising electric field. We find that a non-thermal population represented by a kappa function with a kappa parameter of around 6 leads to good agreement with observations. This value of kappa is slightly lower than the one inferred by \citet{cranmer_suprathermal_2014}. \\

PSP detected surprisingly little wind near the Sun with speeds in excess of 500 km/s
\citep[see, e.g.,][]{rivera_2024,samara_2024}, and additional effects not accounted for in the present study contribute to the acceleration of the fast solar wind to beyond 600 km/s in the inner heliosphere \citep{rivera_2024}. We note that additional sources of momentum could be imparted to protons by other mechanisms, such as reconnection outflows above transient events such as brightpoints known to occur at the base of the solar corona \citep{griton_2020,Gannouni2023, Hou2024}. These reconnection events have been associated with the occurrence of jets and microstreams in the solar wind \citep{Gannouni2023} as well as potential contributors to additional ionisation of heavy ions \citep{Hou2024}. A future study could attempt to run the ISAM model along time-evolving magnetic field lines (i.e. after reconnection events) as ISAM is able to model non-inertial effects. This is done regularly to model ionospheric convection with the ionospheric version of the model employed here \citep{marchaudon_new_2015}. Another source of proton acceleration could also come from momentum exchange with alpha particles, which are known to be differentially heated in the source of the fast solar wind \citep{stansby_2019}. As alpha particles are an important species, they also modify the electron/temperature scale heights and possibly the ionisation rates of heavy ions along the flux tubes. These aspects are the subject of an ongoing study that will be submitted for publication in the coming months. 
\section*{Acknowledgments}
We thank the referee for his constructive comments, which helped improve the quality of this work.


\appendix

\section{Charge state calculation method}
\label{ax:ko_ioniz}
As ions are transported through the solar atmosphere, they undergo ionization and recombination where collisions with electrons are significant. In a stationary hypothesis and in the absence of differential flows between different ionization states within an element, we can combine the conservation equations associated to each ionization state to get a matrix differential equation describing the radial evolution of the ionic fraction $y_i=n_i/(\Sigma^Z_{i=0}n_i)$ where the subscript $i$ indicates the ionization state of the element considered \citep{ko_empirical_1997}. 
\begin{equation}
    u\frac{\partial y_i}{\partial r}=n_e\left(y_{i-1}C_{i-1}(T_e)-y_i(C_i(T_e)+R_{i-1})(T_e)+y_{i+1}R_i(T_e)\right)
    \label{eq:d_y}
\end{equation}
\begin{equation}
    \Sigma^Z_{i=0} \: y_i=1
\end{equation}
Where $n_e$ is the electron density, $T_e$ the electron temperature, $u$ the single velocity for all the ionization state, $C_i$ the ionization rate (including collisional ionization and autoionization) and $R_i$ the recombination rate (including radiative recombination and dielectronic recombination). The reaction rates are taken from the CHIANTI 10.0.2 Atomic Database \citep{dere_chianti_1997,del_zanna_chianti_2021}. This system of equations is solved using the method described in appendix B.

This method has the advantage of being fast compared to the Runge-Kutta method, but could not be used when the lower boundary condition was situated before the sharp gradients of the transition region. We nevertheless chose this method by setting our boundary condition right after the transition region, where collisions are still important and an ionisation equilibrium can still be considered. To test this approach and the hypothesis of negligible differential flows, we compared the results given by this method for the charge-state ratios $n_{O^{7+}}/n_{O^{6+}}$ with the solution given by a full resolution of Oxygen from the bottom of the chromosphere to the solar wind on a fixed background of protons and electrons using ISAM. This point is discussed in Appendix \ref{ax:oxygen}. The results of the comparison gave no significant difference on $n_{O^{7+}}/n_{O^{6+}}$, indeed, the full coupling of Oxygen have shown no significant differential flows in the region where the charge state evolves.
\\ \\
Looking at equation \ref{eq:d_y}, we see that when collisions become too rare (attained either when the density is very low, the temperature is very high or both) the right-hand side vanishes and the ionic fraction remains invariant along the flow trajectory. This is true only when no differential flows are considered, as a relative velocity between ionization states can alter their relative densities leading to an evolution of the ionic fraction even in the collisionless regime. The vanishing of the right-hand side can also happen when the velocity of the flow is so high that a plasma parcel cannot reach ionization equilibrium while in the collisional part of the corona.  Conversely, a low flow velocity and/or a very high collision rate result in ionization equilibrium.
\\ \\
It is important to note that the velocity $u$ in \ref{eq:d_y} is the velocity of the ion considered. In the case of this parametric study, we ran our model for protons and electrons, not heavy ions. Hence, we could not provide the velocity profile of minor ions, instead we provided the velocity of protons. This is equivalent to say $u_i=u_p$ where $u_p$ is the velocity of protons. This is a reasonable assumption as the region where charge state evolves is still collisional, forcing $u_i \approx u_p$ by friction.   This hypothesis has been done in the majority of the literature solving \ref{eq:d_y} using $T_e$, $n_e$, $u_i$ given by solar wind models. 
\\ \\
The recombination and ionization rates given by CHIANTI are calculated assuming a Maxwellian distribution function but the presence of non-Maxwellian features such as a suprathermal electron population would have the effect of enhancing these rates, thus the charge state ratios. In this case, the method presented here has to be adapted. This point is discussed in section \ref{sec:solu_supra} and Appendix \ref{ax:kappa_ioniz}.
\section{Resolution method for the ionic fraction evolution}
\label{ax:2}
Equation \ref{eq:d_y} is of the following form :
\begin{equation}
\frac{d y_i}{dr} = A_{ij} y_j
\label{eq:systemdiff}
\end{equation}
Where the matrix $A_{ii}$ is defined as
\begin{equation}
A_{ii} = - \frac{n_e}{u} \left( C_i + R_{i-1} \right)
\end{equation}
\begin{equation}
A_{i,i-1} =  \frac{n_e}{u} R_i 
\end{equation}
\begin{equation}
A_{i,i+1} =  \frac{n_e}{u} C_{i-1}
\end{equation}
The general solution of this equation is :
\begin{equation}
     y_i(r) = y_i(r_0) \exp(B_{ji}) 
     \label{eq:y_i_sol}
\end{equation}
where $r_0$ is the height of the boundary condition $y_i(r_0)$, $B_{ji}=\int_{r_0}^{r} A_{ji}(r’) \, dr’ $.  The integral $B_{ji}$ is computed numerically using the trapezoidal rule. We then diagonalize numerically $B_{ji} = V_{ji} D_{ii} (V^{-1})_{ji}$. Where $V_{ji}$ and $D_{ii}$ are respectively the matrix of eigenvectors and the diagonal matrix of eigenvalues. With $D_{ii}$ being diagonal, we can write $\exp\left( \int_{r_0}^{r} A_{ji}(r’) \, dr’ \right)=V_{ji} \text{exp}(D_{ii}) (V^{-1})_{ji}$. Where
\begin{equation}
      \text{exp}(D_{ii})=\text{diag}\left(e^{\lambda_1},e^{\lambda_2},....,e^{\lambda_n}\right)
      \label{eq:expD}
\end{equation}
With $\lambda_1,\lambda_2,....,\lambda_n$, the eigenvalues of $A_{ij}$.
\\
The solution is then computed by solving \ref{eq:y_i_sol} using \ref{eq:expD} and taking $r_0$ to be right after the transition region.

\section{Ionisation/Recombination rates for a kappa distribution}
\label{ax:kappa_ioniz}

The sum of N Maxwellian distribution functions each characterized by a different density and temperature.
\begin{equation}
    f_{M}^{\Sigma}(v_e,T_e^1,T_e^2,.....T_e^N,n_e^1,n_e^2,....n_e^N)=\sum_{j=1}^N n_e^j\left(\frac{m_e}{2\pi k_B T_e^j}\right)^{\frac{3}{2}}e^{\left(\frac{m_ev_e^2}{2k_BT_e^j}\right)}
\end{equation}
We choose the following error function
\begin{equation}
    \epsilon_{err}=\sum_{v_e} \left(|\ln{f_\kappa}| - |\ln{f_{M}^{\Sigma}}|\right)^2
\end{equation}

\begin{figure}[!h]
      \centering
      \includegraphics[scale=0.28]{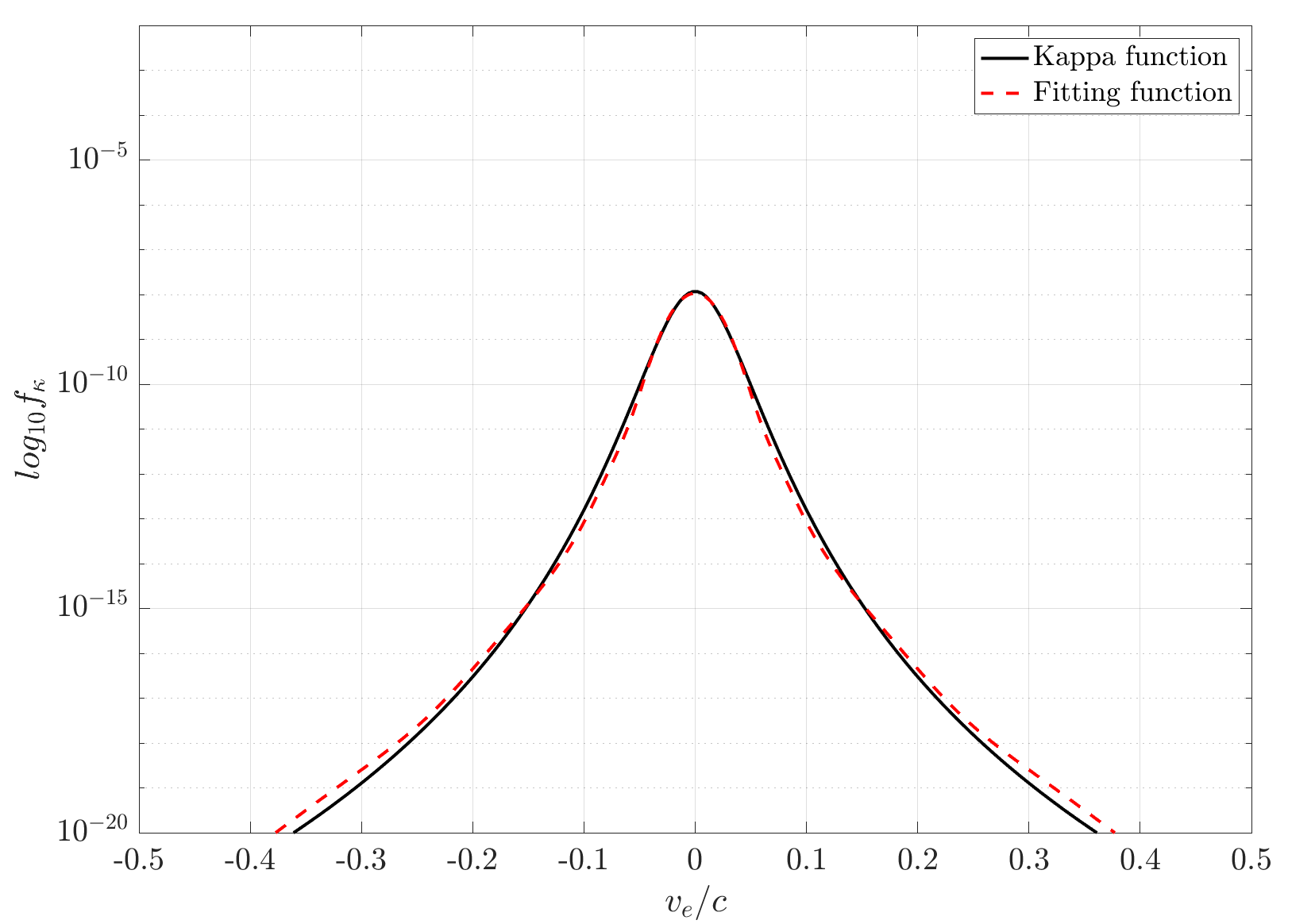}
          \caption{\emph{\ Fit of a kappa function with $\kappa=6$ by 8 partial Maxwellians.} The black curve corresponds to the kappa function. The red curve corresponds to the sum of the 8 partials Maxwellians.
      }
      \label{fig:kappafit}
  \end{figure}

Fitting $f_\kappa$ relies on the minimization of the error function $\epsilon_{err}$ by finding the optimal combination of parameters ($T_e^1,T_e^2,.....T_e^N,n_e^1,n_e^2,....n_e^N$) for each Maxwellian distribution.
The best fit of $f_\kappa$ by $f_{M}^{\Sigma}$ is determined using the minimisation algorithm of the \textit{fminsearch} function in \textit{MATLAB}. We proceed to this minimisation for each grid point of our profiles.  The result of the fit is given in Figure \ref{fig:kappafit} for $N=8$, $\kappa=6$ at the altitude $r=0.1R_\odot$.
\\
The total ionization and recombination rates are then given by :
\begin{equation}
    C=\frac{1}{n_e}\sum_{j=1}^N n_e^j \, C_j(T_e^j)
\end{equation}
\begin{equation}
    R=\frac{1}{n_e}\sum_{j=1}^N n_e^j \, R_j(T_e^j)
\end{equation}
The ionic fraction is then calculated by solving \ref{eq:d_y} using the new ionisation and recombination rates. We tried different $\kappa$ values and their associated increase in charge state ratio $n_{O^{7+}}/n_{O^{6+}}$ and we have determined that $\kappa=6$ is the most adapted choice, which is consistent with the profiles obtained when the ambipolar electric field induced by a Kappa eVDF with $\kappa=6$ is included (Section \ref{sec:solu_supra}).

\section{Solving for oxygen ions}
\label{ax:oxygen}
\begin{figure}[!h]
      \centering
      \includegraphics[scale=0.33]{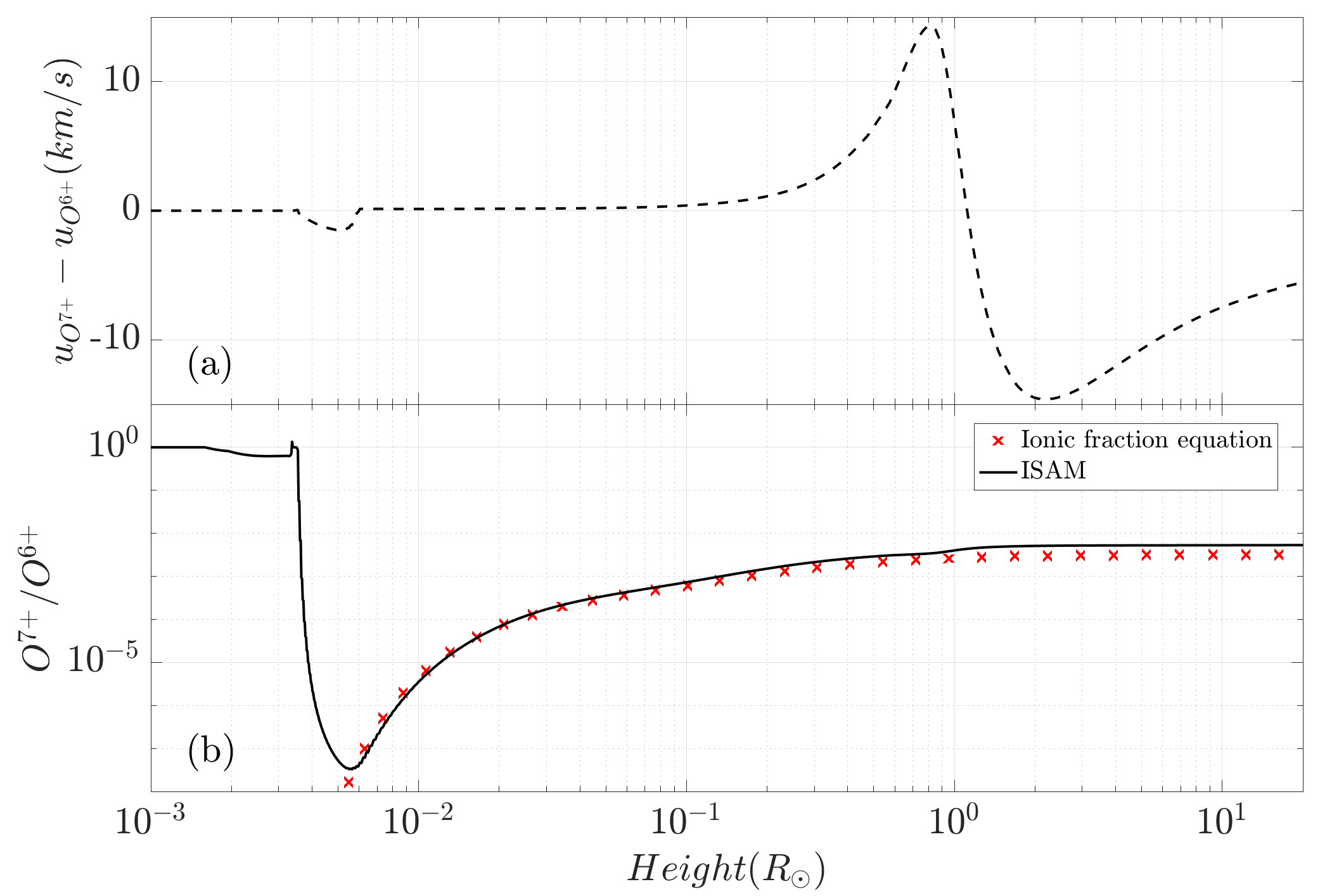}
          \caption{\emph{\ Results of a self-consistent modelisation of oxygen using ISAM.}Panel (a) : $u_{O^{7+}}-u_{O^{6+}}$ calculated from our model (dashed line). Panel (b) : Comparison of $n_{O^{7+}}/n_{O^{6+}}$ from our model (black line) and from solving equation \ref{eq:d_y} (red cross).}
      \label{fig:oxygen}
  \end{figure}

A self-consistent simulation of oxygen and all its ionization states was performed to demonstrate that the method described in Appendix \ref{ax:ko_ioniz} was an appropriate means of modelling the ionic fraction evolution. The modelling of oxygen with ISAM alleviates the hypothesis of equation \ref{eq:d_y} by allowing for differential velocities between ions. Because of the numerical difficulty of a self-consistent simulation of e, H, O and all their ions, we simply simulate a wind for e,H,p and fix it when convergence is reached. On this fixed background, we then solve for oxygen and all its ions $O,O^+,O^{2+},......O^{8+}$. We justify this procedure by the small impact that the oxygen fluid has on the dynamics of hydrogen because of its very low density in the solar wind. This allows us to reduce the simulation time considerably.
\\
Figure \ref{fig:oxygen}, panel (b), illustrates that the discrepancy between the modeled ratio of oxygen  $n_{O^{7+}}/n_{O^{6+}}$, using ISAM and the solution of equation \ref{eq:d_y} is low. At approximately 1 $R_\odot$, the discrepancy is more pronounced and can be attributed to an increase in the differential flow between $O^{6+}$ and $O^{6+}$ ions. Despite the generation of differential flows, we conclude that the difference between the two methods is sufficiently low to justify the use of the CSRs modeled by solving equation \ref{eq:d_y}.


%



\clearpage
\bibliographystyle{aasjournal}
\bibliography{sample631}{}



\end{document}